\documentclass[12pt]{article}

\usepackage{amsmath,amsthm, amsfonts, amssymb, amsxtra, amsopn}
\usepackage{pgfplots}
\pgfplotsset{compat=1.13}
\usepackage{pgfplotstable}
\usepackage{graphicx,grffile}
\usepackage{multirow}
\usepackage{booktabs}
\usepackage{listings}
\usepackage{cmap}
\usepackage{colortbl}
\usepackage{adjustbox}
\usepackage{epsfig}
\usepackage{mdframed}
\usepackage{tabularx}
\usepackage{longtable}

\usepackage[tableposition=top,font=small,skip=5pt]{caption}
\usepackage{subcaption}
\usepackage{makecell} 
\usepackage[explicit]{titlesec}

\PassOptionsToPackage{hyphens}{url}

\usepackage{hyperref}
\hypersetup{colorlinks=true,linkcolor=black,citecolor=black,urlcolor=blue,filecolor=black}
\hypersetup{pdfpagemode=UseNone,pdfstartview=}

\usepackage{enumitem}
\setlist[itemize]{noitemsep, topsep=0pt}

\advance\oddsidemargin by -0.35in
\advance\textwidth by 0.7in

\advance\topmargin by -0.4in
\advance\textheight by 0.8in

\long\def\symbolfootnotetext[#1]#2{\begingroup%
\def\thefootnote{\fnsymbol{footnote}}\footnotetext[#1]{#2}\endgroup}

\newcommand\dunderline[3][-1pt]{{%
  \sbox0{#3}%
  \ooalign{\copy0\cr\rule[\dimexpr#1-#2\relax]{\wd0}{#2}}}}
\def\uuu{\kern-1pt\dunderline{0.75pt}{\phantom{M}}}

\def\z{{\tt z}}

\def\zz{\phantom{0}}

\title{Image-Based Malware Classification Using QR and Aztec Codes}

\author{Atharva Khadilkar\footnotemark[1]\ \ \ 
Mark Stamp\footnotemark[1]\,\,\footnotemark[2]}

\begin{document}

\symbolfootnotetext[1]{Department of Computer Science, San Jose State University}
\symbolfootnotetext[2]{mark.stamp$@$sjsu.edu}

\maketitle

\abstract
In recent years, the use of image-based techniques for malware detection has gained prominence, 
with numerous studies demonstrating the efficacy of deep learning approaches such as Convolutional Neural Networks (CNN) 
in classifying images derived from executable files. In this paper, we consider an innovative method that relies on an image 
conversion process that consists of transforming features extracted from executable files into QR and Aztec codes. 
These codes capture structural patterns in a format that may enhance the learning capabilities of CNNs. 
We design and implement CNN architectures tailored to the unique properties of these codes and apply 
them to a comprehensive analysis involving two extensive malware datasets, both of which include a significant corpus of 
benign samples. Our results yield a split decision, with CNNs trained on QR and Aztec codes outperforming the
state of the art on one of the datasets, but underperforming more typical techniques on the other dataset.
These results indicate that the use of QR and Aztec codes as a form of feature engineering
holds considerable promise in the malware domain, and that additional research is needed
to better understand the relative strengths and weaknesses of such an approach.

\section{Introduction}

In the current digital age, cybersecurity threats have become increasingly sophisticated, one example of which is obfuscated malware. 
Obfuscated malware is malware that is ``disguised'' so that it is difficult to detect using conventional methods. 
Traditional antivirus systems rely on signature-based detection, 
which struggles to identify obfuscated malware~\cite{Aycock}. 

Malware classification is the process of categorizing various types of malware into distinct groups based on their behavior, 
characteristics, or potential impact. The conventional approach to malware classification is reliant on signature based 
and heuristic methods. In the malware context, signatures typically consist of known patterns that appear in the code, 
whereas heuristic analysis usually attempts to focus on behavioral patterns. These techniques struggle with the obfuscation 
strategies employed in modern malware, often leading to false positives or negatives.

Traditional machine learning (ML) methods have been used for enhancing malware classification offering an alternative 
to conventional signature based approaches~\cite{Stamp18}. A variety of traditional methods are commonly employed, including 
logistic regression, Support Vector Machine (SVM), and Random Forests. ML techniques significantly improves the 
detection of novel malware strain, as ML does not solely rely on pre-existing signatures. However, such ML methods 
can be challenged by sophisticated obfuscation techniques. A recent trend in malware detection consists of converting 
executable files to images and using sophisticated image analysis techniques for classification. This approach is promising 
and has shown improved results over traditional ML techniques. However, the process used to convert executables to 
images can have a large impact on the success of such image-based analysis techniques.

Building on traditional ML methods for malware classification, we propose to use QR and Aztec codes as image 
representations of data, combined with advanced image classification techniques. By leveraging the unique patterns 
within these images, our method aims to improve on image-based analysis of obfuscated malware. We find that our approach 
% more robust?
offers an improvement over traditional features and also improves on typical image-generation techniques
for the detection of complex malware variants.

The remainder of this paper is organized as follows. Section~\ref{chap:RW} discusses relevant previous work, including traditional 
and image-based techniques for malware detection and classification. Section~\ref{chap:DS} introduces the datasets 
utilized in this study, detailing their composition, source, and the methodology employed for their collection and preparation. 
Section~\ref{chap:LIB} outlines the libraries and platforms used in our experiments. Following this, Section~\ref{chap:METHOD} 
covers the  techniques and methodologies employed, along with a discussion of the implementation of these methods. 
Section~\ref{chap:RESULTS} includes the results of our experiments and compares our results to previous 
related work. Section~\ref{chap:CONCLUSION} concludes the paper, summarizing our key findings and discussing potential 
avenues for future research.

\section{Selected Related Work}\label{chap:RW}

This section explores selected prior research in the realm of malware detection, particularly focusing on approaches that use 
image representation for identifying malware. Additionally, we explore studies related to the detection of obfuscated malware, 
emphasizing techniques that utilize memory dumps. These topics represent significant advances in malware detection.

\subsection{Obfuscated Malware}

Obfuscated malware techniques complicate the detection process by disguising the malicious code, making it challenging 
for traditional antivirus solutions to identify threats effectively. Techniques such as polymorphism and metamorphism are 
frequently employed, allowing malware to alter its code with each replication, thereby evading signature-based detection 
systems. The research in~\cite{5633410} detail these methods, noting their sophistication and the difficulty they pose to 
detection efforts. Similarly, \cite{10.1145/2160158.2160159} discuss the theoretical underpinnings of code obfuscation, 
pointing out the effectiveness of such techniques in protecting malware from analysis. These approaches highlight the 
continuous arms race between cybersecurity professionals and attackers, underscoring the need for advanced detection 
methods capable of penetrating these obfuscations.

The CIC-MalMem-2022 dataset~\cite{malmem} 
contains features extracted from obfuscated malware samples. Studies focusing on binary classification using 
the CIC-MalMem-2022 dataset, such as~\cite{dener2022malware,a15090332}, have applied various ML techniques, 
achieving up to~0.9997 accuracy with learning methods such as Decision Trees and SVC. Additionally the research 
in~\cite{a15090332} employs feature engineering and tree-based techniques such as XGBoost and CatBoost, 
achieving a~1.00 accuracy in binary classification with a Random Forest Classifier. This demonstrates the effectiveness 
of combining advanced methods and feature engineering in improving malware detection.

\subsection{Behavioral Analysis of Malware}

Malware behavior analysis techniques focus on observing the actions of malware within a system, offering insights beyond 
those that static analysis can provide. This dynamic approach, as considered in~\cite{bayer2006dynamic} and~\cite{SHABTAI200916},
for example, involves monitoring the execution patterns and network behaviors of malware to classify and understand its nature. 
Such techniques are critical in identifying new variants of malware by examining their behavior patterns, potentially offering more 
adaptive and robust mechanism for threat detection. The effectiveness of these methods lies in their ability to provide a detailed 
view of malware operations, contributing to the development of more precise and effective cybersecurity measures.

\subsection{Image Representations}

The conversion of malware binaries into visual images for analysis has recently shown great promise. 
The papers~\cite{7492653,10.1145/2016904.2016908,Prajapati2021} explore the potential of such techniques, 
which leverages the visual patterns that emerge from the binary code of malware when represented as images. These methods 
allow for the application of advanced image processing techniques to identify distinctive features associated with malicious software. 
The advantage of image representation lies in its ability to reveal patterns that are not easily discernible through traditional binary analysis, 
providing an alternative avenue for the detection and classification of malware.

The use of Convolutional Neural Networks (CNNs) for classifying malware based on image representations 
showcases the application of deep learning in cybersecurity. For example, the research in~\cite{8328749,7492653} 
demonstrates the effectiveness of CNNs in analyzing the visual patterns of malware images to accurately classify different types of 
malware. These studies highlight the ability of CNNs to learn and identify complex patterns within images, facilitating a highly 
effective classification system. Another strength of this approach lies in its capacity to process and analyze large datasets of 
malware images, thus offering a scalable and efficient solution for the identification of malware.

\subsection{Advanced CNN Architectures}

The convergence of Convolutional Neural Networks and pre-trained models has significantly advanced malware classification, 
leveraging deep learning for cybersecurity. The studies in~\cite{app14062614, panda2023ensemble}, for example, 
highlight the effectiveness of pre-trained CNN models. Examples of such pre-trained models include VGG16, ResNet-50, and DenseNet-201. 
Additionally, these works delve into the challenges associated with feature extraction, feature engineering, computational demands, 
dataset imbalances, and so on. These are issues often encountered when learning techniques are used in the malware domain.

Further innovations include ensemble and transfer learning approaches to enhance accuracy and efficiency in malware-related tasks. 
For instance, in~\cite{app11146446,panda2023ensemble} the application of fine-tuned CNN-based transfer learning models on 
transformed 2D images of malware binaries demonstrated exceptional detection accuracies, outperforming conventional 
methods. More generally, these approaches signify a shift towards utilizing deep learning models to deal with the challenges of 
evolving cybersecurity threats.

%The study in~\cite{R2018255} explores using Deep Neural Networks (DNN) for malware classification, 
%developing a DNN model trained on the EMBER dataset, showing an accuracy rate of~0.989. 

\subsection{Memory Dump Analysis}

Memory dump analysis for malware classification involves examining snapshots of system memory to detect potentially
malicious behavior. This technique, as discussed in~\cite{CASE201723,casey2009handbook,Chakkaravarthy2019}
provides valuable insights into the runtime behavior of malware. Memory dump analysis is particularly effective in identifying malware 
that employs evasion techniques, as it allows for the examination of the system state at the time of execution, at which point most
obfuscation techniques have run their course. This approach enhances the capability to classify and analyze complex malware,
underscoring its importance in the comprehensive examination of cyber threats.

\subsection{QR Codes} 

Quick Response (QR) codes are 2D bar codes that can encode virtually any type of data, and
are easily readable by devices such as smartphone cameras. A QR code can encode
up to~7089 digits, 4296 alphanumeric characters, 2953 bytes, or~1817 Kanji characters,
although these values may be reduced, depending on the level of error correction that is
applied~\cite{qr}.

For our purposes, QR codes provide a basis for applying image-based learning techniques to 
virtually any type of data. The research in~\cite{a16030160} exemplifies this by evaluating 
various pre-trained CNN models, including AlexNet and MobileNetv2, to accurately identify the source printer of QR codes. 
This example highlights the potential for merging QR code versatility with CNN image-based learning capabilities, 
offering a novel pathway for data classification that has the potential to significantly enhance information security.
The paper~\cite{10.1145/3326459.3329166} explores another approach which uses CNNs to classify malware based 
on QR code representations of data. 

\subsection{Aztec Codes}

Aztec codes are similar to QR codes, but they can be more space-efficient
for a given amount of data.
An Aztec code can encode a maximum of~3832 numeric digits,
3067 alphabetic characters, or 1914 bytes of data~\cite{aztec}.
% Add more here?
As far as the authors are aware, Aztec codes have
not been previously studied in the context of malware analysis,
or in conjunction with image-based learning techniques.

\section{Datasets}\label{chap:DS}

In this section, we provide an overview of the datasets used in this research. We consider two distinct datasets, 
one of which consists of dynamic features extracted from obfuscated malware, while the other consists of
static features extracted from typical malware.

\subsection{CIC-MalMem-2022}

The paper~\cite{carrier2022detecting} focuses on improving malware detection, specifically targeting obfuscated malware. 
The study uses the VolMemLyzer tool, a memory feature extractor, to better identify hidden and obfuscated malware. 
A significant contribution is the creation of the MalMemAnalysis-2022 dataset, which includes over~2500 malware samples 
in the broad categories of spyware, ransomware, and Trojan horse, as well as a representative benign set. 
The authors employ a stacked ensemble ML model for detection, achieving high 
accuracy and F1-score for the binary classification problem.

Based on the work in~\cite{carrier2022detecting}, the CIC-MalMem-2022 dataset was published~\cite{malmem}. 
This dataset includes features
extracted using a memory dump operation in debug mode. This method is specifically designed to prevent the dumping process 
itself from being recorded in the memory dumps, which ensures that only the relevant data is captured.
The dataset consists of a total of~58,596 samples extracted from~2916 malware executables and~2916 benign executables, with
a minimum of~100 and a maximum of~200 samples per executable.
The malware executables are from three major categories, namely, ransomware, spyware, and Trojan horse.

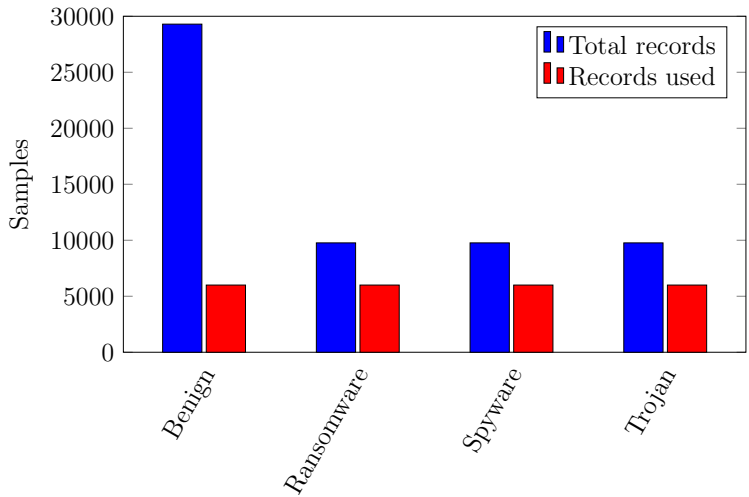
\begin{figure}[!htb]
    \centering
    \begin{tikzpicture}[scale=0.825, every node/.style={scale=1.0}]
\pgfkeys{/pgf/number format/.cd,1000 sep={}}
\begin{axis}[%bar shift=0pt,
        width  = 0.75*\textwidth,
        height = 7.0cm,
        ymin=0, ymax=30000,
        ytick={0, 5000, 10000, 15000, 20000, 25000, 30000},
        ytick scale label code/.code={},
        major x tick style = transparent,
        ybar=5*\pgflinewidth,
        bar width=18pt,
%        ymajorgrids = true,
%        xlabel = {Learning technique},
        ylabel = {Samples},
        symbolic x coords={Benign, Spyware, Ransomware, Trojan},
        xticklabels={Benign, Spyware, Ransomware, Trojan},
        yticklabels={0, 5000, 10000, 15000, 20000, 25000, 30000},
	y tick label style={
%%		rotate=60,
%    		/pgf/number format/.cd,
%   		fixed,
%   		fixed zerofill,
%%		sep=,
%    		precision=0
	},
%	yticklabel pos=right,
        xtick = data,
        x tick label style={
        		rotate=60,
		yshift=5pt,
		xshift=2pt,
		anchor=north east,
		},
%		font=\small},
%        scaled y ticks = false,
	%%%%% numbers on bars and rotated
%        nodes near coords,
%        every node near coord/.append style={rotate=90, 
%        								   anchor=west,
%								   %font=\footnotesize,
%								   /pgf/number format/.cd,
%								   fixed,
%								   fixed zerofill,
%%								   sep=,
%								   precision=4},
        %%%%%
%        enlarge x limits=0.03,
%        enlarge x limits=0.06,
        enlarge x limits=0.175,
        legend cell align=left,
        legend pos=north east,
%        every axis plot/.append style={
%        		bar shift=0pt,
%		fill
%        },
%        legend style={
%%                at={(1,1.05)},
%%                anchor=south east,
%%	        nodes={rotate=90},%%%%% rotate text in legend
%%                at={(0.125,0)},
%%                at={(0.125,0)},
%%                at={(0.8775,0)},
%                at={(0.82,0.56)},
%                anchor=south,
%                column sep=1ex
%        },
]
\addplot [fill=blue,opacity=1.00]
coordinates {
(Benign, 29298)
(Ransomware, 9766)
(Spyware, 9766)
(Trojan, 9766)
};
\addlegendentry{Total records}
\addplot [fill=red,opacity=1.00]
coordinates {
(Benign, 6000)
(Ransomware, 6000)
(Spyware, 6000)
(Trojan, 6000)
};
\addlegendentry{Records used}
\end{axis}
\end{tikzpicture}
    \caption{CIC-MalMem-2022 class distribution}\label{fig: MalMem CIC Initial distribution}
\end{figure}

From this dataset, we use~6000 samples chosen randomly for each of the three malware categories,
and~6000 benign samples. These~24,000 samples are the basis for training and testing our multiclass models 
using the learning techniques discussed in Section~\ref{chap:METHOD}, below.
The distribution of samples from the CIC-Malmem-2022 dataset 
is given in Figure~\ref{fig: MalMem CIC Initial distribution}.

\subsection{BODMAS}

The BODMAS dataset~\cite{yang2021bodmas} is a collaborative effort between Blue Hexagon 
and the University of Illinois at Urbana-Champaign (UIUC), and it represents a valuable resource for the cybersecurity research
community. This dataset consists of~57,293 malware samples and~77,142 benign samples, collected between 
August~2019 and September~2020. This substantial collection, which includes samples spanning~581 malware families, 
is notable for including date of origin of each sample, thus providing a resource for temporal-based analysis and classification 
of malware.

The feature vectors for each of the malware sample was extracted using the LIEF project, 
similar to the EMBER~\cite{anderson2018ember} dataset. Executable file formats share common features 
including symbols, relocations, and entry-point. Each malware sample has a feature vector of dimension~2384 
and associated metadata. This metadata includes details such as timestamp, label (malware or benign), 
and the specific malware family of the sample. The feature vectors consist of features parsed from the PE file, 
including the SHA256 hash of the file, header characteristics, entry points, entropy, and various 
histograms~\cite{anderson2018ember}.

\begin{figure}[!htb]
    \centering
    \begin{tikzpicture}[scale=0.825, every node/.style={scale=1.0}]
\pgfkeys{/pgf/number format/.cd,1000 sep={}}
\begin{axis}[%bar shift=0pt,
        width  = 0.95*\textwidth,
        height = 8.0cm,
        ymin=0,ymax=5300,
        ytick={0, 1000, 2000, 3000, 4000, 5000},
        major x tick style = transparent,
        ybar=5*\pgflinewidth,
        bar width=11.5pt,
%        ymajorgrids = true,
%        xlabel = {Learning technique},
        ylabel = {Samples},
        symbolic x coords={Benign, Sfone, Wacatac, Upatre, Wabot, Small, Ganelp, Dinwod, Mira, Berbew, 
        		Sillyp2p, Ceeinject, Gepys, Benjamin, Musecador},
        xticklabels={Benign, \texttt{Sfone}, \texttt{Wacatac}, \texttt{Upatre}, \texttt{Wabot}, \texttt{Small}, \texttt{Ganelp}, 
        		\texttt{Dinwod}, \texttt{Mira}, \texttt{Berbew}, \texttt{Sillyp2p}, \texttt{Ceeinject}, \texttt{Gepys}, \texttt{Benjamin}, \texttt{Musecador}},
	y tick label style={
    		/pgf/number format/.cd,
   		fixed,
   		fixed zerofill,
%		sep=,
    		precision=0},
%	yticklabel pos=right,
        xtick = data,
        x tick label style={
        		rotate=60,
		yshift=5pt,
		xshift=2pt,
		anchor=north east,
		},
%		font=\small},
%        scaled y ticks = false,
	%%%%% numbers on bars and rotated
%        nodes near coords,
%        every node near coord/.append style={rotate=90, 
%        								   anchor=west,
%								   %font=\footnotesize,
%								   /pgf/number format/.cd,
%								   fixed,
%								   fixed zerofill,
%%								   sep=,
%								   precision=4},
        %%%%%
%        enlarge x limits=0.03,
%        enlarge x limits=0.06,
        enlarge x limits=0.05,
        legend cell align=left,
        legend pos=north east,
        every axis plot/.append style={
        		bar shift=0pt,
		fill
        },
%        legend style={
%%                at={(1,1.05)},
%%                anchor=south east,
%%	        nodes={rotate=90},%%%%% rotate text in legend
%%                at={(0.125,0)},
%%                at={(0.125,0)},
%%                at={(0.8775,0)},
%                at={(0.82,0.56)},
%                anchor=south,
%                column sep=1ex
%        },
]
\addplot [fill=green,opacity=1.00]
coordinates {
(Benign, 5200)
(Sfone, 0)
(Wacatac, 0)
(Upatre, 0)
(Wabot, 0)
(Small, 0)
(Ganelp, 0)
(Dinwod, 0)
(Mira, 0)
(Berbew, 0)
(Sillyp2p, 0)
(Ceeinject, 0)
(Gepys, 0)
(Benjamin, 0)
(Musecador, 0)
};
\addlegendentry{Benign}
\addplot [fill=red,opacity=1.00]
coordinates {
(Benign, 0)
(Sfone, 4729)
(Wacatac, 4694)
(Upatre, 3901)
(Wabot, 0)
(Small, 0)
(Ganelp, 0)
(Dinwod, 0)
(Mira, 0)
(Berbew, 0)
(Sillyp2p, 0)
(Ceeinject, 0)
(Gepys, 0)
(Benjamin, 0)
(Musecador, 0)
};
\addlegendentry{Malware samples used}
\addplot [fill=blue,opacity=1.00]
coordinates {
(Benign, 0)
(Sfone, 0)
(Wacatac, 0)
(Upatre, 0)
(Wabot, 3673)
(Small, 3339)
(Ganelp, 2232)
(Dinwod, 2057)
(Mira, 1960)
(Berbew, 1749)
(Sillyp2p, 1616)
(Ceeinject, 1169)
(Gepys, 1124)
(Benjamin, 1071)
(Musecador, 1054)
};
\addlegendentry{Malware not used}
\end{axis}
\end{tikzpicture}
    \caption{Benign and BODMAS class distribution}\label{fig:BODMAS_DIST_ORIG}
\end{figure}
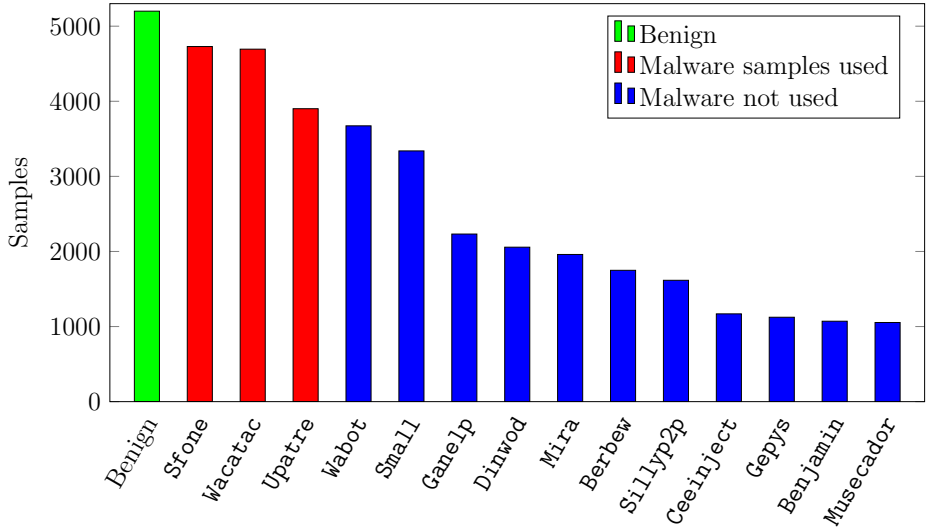

Since the BODMAS dataset consists of approximately~58,000 malware samples from~581 malware families, 
we selected a handful of the most frequently occurring families for our analysis, along with a subset of benign samples.
The distributions of the benign class and the
%%%%% The numbers that were here did not match the info for the BODMAS dataset ?????
top~10 families in the BODMAS dataset are shown using Figure~\ref{fig:BODMAS_DIST_ORIG}.
We have selected~13,324 samples consisting of the top three malware families, namely, 
\texttt{Sfone}, \texttt{Wacatac}, and \texttt{Upatre}, %and \texttt{Wabot} 
along with~5200 benign samples for a total of~18,524 samples. Similar to the CIC-MalMem-2022 dataset, 
we apply the learning techniques discussed in Section~\ref{chap:METHOD}, below.

\section{Implementation}\label{chap:LIB}

This section provides a comprehensive overview of the software and libraries used in our research. 
Here, we introduce each chosen platform and library, specifying its role, benefits, and the reason for 
its selection.

\subsection{Machine Learning Tools}

The implementation of our research involves data exploration of two datasets. 
Along with this, we also explores feature selection and model training. The models vary 
from classical ML to pre-trained DL models.
These models varying greatly in the amount of complexity, and they required different libraries to be used. 
Due to the availability of necessary packages, Python~3.9 was used throughout our experiments. 
Specifically, the following Python libraries were used.

\begin{itemize}
\item{\textbf{scikit-learn}} ---
The Python library \texttt{scikit-learn} is designed for machine learning~\cite{scikit-learn}. It offers tools for data preprocessing, 
model building, and evaluation. It also includes algorithms for both supervised and unsupervised learning 
such as regression, decision trees, clustering, and Support Vector Machines. \texttt{scikit-learn} is designed to work with 
NumPy and SciPy. We use \texttt{scikit-learn} for some of the classical ML algorithms and for feature selection. 
\texttt{scikit-learn} was also used to calculate our performance metrics (accuracy, F1-score) for our 
classic ML models.

\item{\textbf{TensorFlow}} ---
TensorFlow is an open-source library for numerical computation and ML~\cite{tensorflow}. It provides a flexible 
platform for building and deploying a wide range of ML models. TensorFlow supports deep learning algorithms 
along with many traditional ML models. The library includes tools for data processing, model creation, training, and inference.
TensorFlow was used to form our generated QR and Aztec codes into image datasets. It was also used to construct and 
train our Convolutional Neural Networks.
\end{itemize}

%\subsection{Visualization}
%
%The following libraries were used for the visualizations used in this research.
%
%\begin{itemize} 
%\item{\textbf{Matplotlib}}
%The module pyplot from Matplotlib was primarily used for 2D visualizations~\cite{matplotlib}. 
%It provides a MATLAB-like interface for creating figures and graphs. Users can generate plots, 
%histograms, power spectra, bar charts, error charts, scatterplots, etc. It is widely used for data 
%visualization in scientific computing and has tools for customizing plots with labels, grids, and various types of lines or markers.
%
%\item{\textbf{Seaborn}}
%Seaborn is a Python data visualization library based on matplotlib~\cite{seaborn}. It offers a high-level interface for 
%drawing informative statistical graphics. Seaborn simplifies the process of creating visualizations from data stored 
%in Pandas DataFrames. It includes functions for creating a variety of plot types, including scatter plots, line plots, 
%histograms, and heatmaps. It was primarily used to generate the confusion matrices for the generated classification reports.
%\end{itemize}

\subsection{Utilities}

The following packages and libraries were also used in our research.
These tools are not directly used for ML, but they are necessary to prepare
the data for our models.

\begin{itemize}
\item{\textbf{Pandas}} ---
Pandas is a Python library for data manipulation and analysis~\cite{pandas}. It provides data structures including
DataFrame and Series for handling tabular data. Pandas is equipped with tools for reading and writing data between 
in-memory data structures and different file formats. We use this library for the manipulation and preprocessing of our data.

\item{\textbf{NumPy}} ---
NumPy is a fundamental package for scientific computing in Python~\cite{numpy}. It provides support for large
arrays and matrices, along with a collection of mathematical functions to operate on these arrays. 
NumPy was used along with Pandas to enabling efficient processing of the data.

\item{\textbf{Qrcode}} ---
The Python library \texttt{qrcode} is designed to generate QR codes~\cite{qrcode}. It allows for the creation and 
customization of QR codes that can encode a wide range of data types, including URLs, text, or numerical information. 
This library provides a simple interface for QR code generation, offering flexibility in terms of size, border, and 
error correction levels. Of course, we employ this library to create QR representations of our data.

\item\textbf{AztecCode} ---
AztecCode from \texttt{aztec\_code\_generator} is a Python library designed for creating Aztec codes, a type of 2D 
barcode that can store a significant amount of data within a small space~\cite{azteccode}. Similar to QR codes, 
but with significant differences in design and capacity, Aztec codes are used in various applications, especially 
where space and readability are critical. The \texttt{aztec\_code\_generator} library provides functionality 
to generate and customize these codes, including setting size, encoding data, and adjusting error correction 
levels. This library was used to generate the Aztec code representation of the data used in our experiments.

\item{\textbf{Operating System}} ---
The standard Python library \texttt{OS} provides a way to use operating system-dependent 
functionality~\cite{pythonos}. It includes functions for interacting with the file system, such as creating, 
listing, and deleting files and directories. We primarily use OS for data organization during the image 
generation process.

\item{\textbf{Pillow}} ---
The \texttt{Image} module from Pillow, the Python Imaging Library, supports opening, manipulating, and saving many different 
image file formats~\cite{pillow}. It provides a wide array of image processing capabilities, including image transformations 
(e.g., rotation and scaling), filtering, enhancement, and so on. We use
Pillow for post processing on our generated QR and Aztec codes.
\end{itemize}

\subsection{Development Platforms}

Google Colab was an essential part of our model training. Here, we provide some details on our
use of Colab and also our local computer setup.

\begin{itemize}
\item{\textbf{Google Colab}} ---
Throughout our experiments, the Google Colab platform was used extensively for data processing, 
data conversion, and training learning models. The platform provides access to most of the libraries
discussed above, including TensorFlow and \texttt{scikit-learn}. The platform offers multiple environment runtimes 
with the option of choosing GPUs. Using GPU hardware acceleration reduced the training times for 
our CNNs by a factor of about four, as compared to the CPU hardware accelerators.

\item{\textbf{Local Computer}} ---
In addition to Google Colab, we used a local desktop setup to execute some of our experiments. 
The local setup served multiple purposes, including visualization and data exploration, as well as
for data preprocessing and cleaning. We also generated the QR and Aztec code images 
using this local machine. The operations were performed locally using Visual Studio Code by 
creating a Python~3.9 virtual environment. The specifications of the local machine are
given in Table~\ref{tab:mach}.
\end{itemize}

\begin{table}[!htb]
    \caption{Local machine}\label{tab:mach}
    \centering
    \begin{tabular}{c|c}
    \toprule
    System & Specification  \\ \midrule
    CPU & AMD Ryzen 5600X \\
    CPU clock rate & 4.6 GHz \\
    GPU & NVIDIA RTX 3080 (10GB) \\
    CUDA core & 8704 \\
    Cores & 6 \\
    RAM & 32GB \\
    OS & MS Windows 10 \\
    \bottomrule
    \end{tabular}
\end{table}

%\begin{itemize}
%    \item CPU: AMD Ryzen 5600X
%    \item GPU: NVIDIA RTX 3080 (10GB)
%    \item RAM: 32GB
%    \item OS: MS Windows 10
%\end{itemize}
%\end{itemize}

\section{Methodology}\label{chap:METHOD}

In this section we describe the machine learning models and methods used to generate our experimental 
results. This section introduces each learning method and provides reasons why
these methods were used in the context of this research.

\subsection{Feature Selection}

Considering the large number of features in our datasets, feature reduction is an important 
aspect to this research. For this purpose we consider the following feature selection methods.

\texttt{SelectKBest} from \texttt{scikit-learn} is a statistical method used to select features that have the most significant 
relationship with the output variable. This works by applying a chosen statistical test to each feature to determine
 its strength of association with the output variable. The \texttt{K} in \texttt{SelectKBest} refers to the number of
 features to select based, on their ranking. We selected features after normalizing using the standard scaler.

Statistical tests available in \texttt{SelectKBest} include the ANOVA F-test for continuous data and~$\chi^2$ for categorical data. 
We have used~$\chi^2$ for selecting our $K$ best features. This method is effective in feature reduction, helping to improve 
model performance by eliminating irrelevant or redundant features. It is particularly useful in helping us reduce the number of
% How much ? 
features before converting the data into a QR or Aztec code, since these representation can only hold a limited number of bytes.

%Principal Component Analysis (PCA) is an algorithm that aims to reduce the dimensionality of a dataset. 
%This reduction is accomplished by transforming the dataset into a set of linearly uncorrelated variables 
%known as principal components. We employed \texttt{scikit-learn} for PCA.

%% How/where, exactly, is PCA used?
%PCA is a powerful tool for analyzing data, reducing dimensionality, simplifying the complexity in high-dimensional data 
%while retaining relevant information. This method is particularly useful for enhancing interpretability but also for preparing 
%datasets for ML models, as it can significantly reduce the time and storage space required for data processing. 
%%PCA achieves this by identifying the axes that maximize the variance of the data and projecting it onto a new 
%%subspace with fewer dimensions.

\subsection{Machine Learning Models}

In this section, we describe all of the ML models used in this research including the classical and deep learning models. 
We also mention why each model was chosen for this research.

\subsubsection{Random Forest}

The Random Forest classifier from \texttt{scikit-learn} is an ML algorithm for classification tasks. It operates by 
constructing multiple decision trees during the training phase and outputs the class that is the mode of the 
classes of the individual trees. This approach to combining multiple models to improve the overall result is 
an example of ensemble learning.

The Random Forest algorithm can handle both numerical and categorical data and is capable of dealing with large 
datasets efficiently. Additionally, it provides measures of feature importance, which can be used for feature selection. 
Random Forest is widely used across various fields for its robustness against overfitting, compared to a single decision tree, 
making it a popular choice for complex classification problems.

\subsubsection{Support Vector Machine}

Support Vector Machine (SVM) from \texttt{scikit-learn} is a supervised ML algorithm used for both classification and 
regression tasks, though it is primarily known for classification. The core principle of SVM is to find the hyperplane 
that best divides a dataset into classes. SVM is distinctive for its use of kernels, which transform the input data space 
into a higher dimensional space where it becomes easier to separate the data linearly. This makes SVM  effective 
for complex datasets where the relationship between features may not be clear. 

The performance of SVMs depend heavily on the selection of the kernel and the tuning of the hyperparameters, 
which can sometimes make it challenging to optimize. SVM is widely utilized in applications ranging from image 
classification to bioinformatics, due to its robustness and versatility.

Support Vector Classifier (SVC) is a generalization of SVM to the multiclass case. SVC
works by finding multiple hyperplanes that best separate different classes with the maximum margin. 
This makes SVC particularly effective for complex classification problems where the decision 
boundary is not immediately obvious.

\subsubsection{Multilayer Perceptron}

The Multilayer Perceptron (MLP) classifier from \texttt{scikit-learn} is a basic type of artificial neural network that is used for 
classification and regression tasks. Unlike simpler linear models, MLP can model complex nonlinear relationships between 
inputs and outputs. An MLP includes at least three layers consisting of an input layer, one or more hidden layers, 
and an output layer. The nodes, or neurons, in each layer are fully connected to those in the next layer, and 
activation functions serve to introduce nonlinearity to the learning process~\cite{Stamp_2022}. 

Training an MLP involves adjusting the weights of the connections through a process known as backpropagation, 
which minimizes the difference between the actual and predicted outputs. MLP is particularly useful for problems 
where the relationship between input and output is not linearly separable. The performance of an MLP is influenced 
by various factors, including the number of hidden layers, the size of these layers, and the choice of activation function. 
MLPs are widely used in a variety of fields, including speech recognition and natural language processing.

\subsubsection{Convolutional Neural Networks}

Convolutional Neural Networks (CNNs) are a basic class of deep neural networks, widely utilized in the field of computer 
vision. Developed with inspiration from the human visual cortex, CNNs excel at automatically and adaptively learning spatial 
hierarchies of features from image data. They consist of multiple layers, including convolutional layers that 
capture patterns such as edges and textures, pooling layers that reduce dimensionality and computational complexity, 
and fully connected layers that classify the images based on the features extracted by convolutional and pooling layers.

Our CNNs are  implemented using the keras library, a high-level neural networks API 
that runs on top of TensorFlow~\cite{tensorflow}. Keras provides a user-friendly interface for 
building and training CNN models, offering a flexible and efficient way to design deep learning models with 
just a few lines of code.

Due to the different features available with the two datasets we consider, 
we employ two distinct CNN architectures. Here, we describe both of these architectures in some detail.

\begin{figure}[!htb]
    \centering
    \begin{tabular}{c}
    \includegraphics[width=75mm]{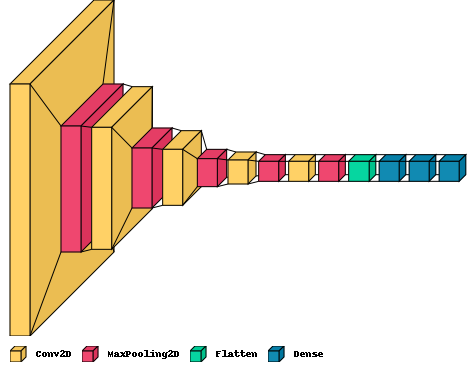}
    \\
    (a) Model for CIC-MalMem-2022 dataset
    \\ \\
    \includegraphics[width=90mm]{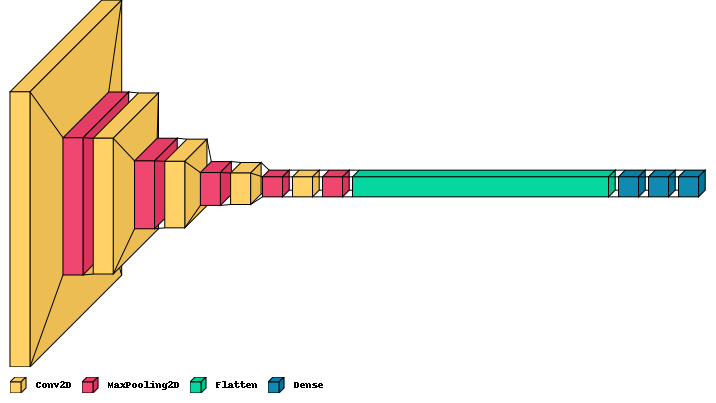}
    \\
    (b) Model for BODMAS dataset
    \end{tabular}
    \caption{CNN architectures}\label{fig:CNN_both}
\end{figure}

\begin{itemize}
\item{\textbf{CNN for CIC-MalMem-2022 dataset}} ---
Our CNN architecture for the CIC-MalMem-2022 dataset
has an input size of~$128 \times 128\times 1$ which is the size of the QR and Aztec codes generated from the dataset.
The initial convolutional layer consists of~32 filters of size~$3\times 3$, which yields an output of size~$126\times 126\times 32$.
There are four more convolutional layers which have output sizes of~$61 \times 61\times 64$, 
$28 \times 28\times 128$, $12 \times 12\times 256$, and~$4 \times 4\times 512$, respectively.
Each convolutional layer is followed by a max pooling layer and the final output of all these
layers is~$2 \times 2 \times 512$. This is then flattened and forwarded to three dense layers for classification.
Our CNN for the CIC-MalMem-2022 dataset is illustrated in Figure~\ref{fig:CNN_both}(a).

\item{\textbf{CNN for BODMAS dataset}} ---
Our CNN for the BODMAS dataset has a similar structure as that 
for the CIC-MalMem-2022 dataset. However, for this CNN the input layer is~$395 \times 395\times 1$,
since this is the size of our QR images generated from the BODMAS data. 
This slightly alters the remaining parts of the model.

The initial convolutional layer consists of~32 filters of size~$3\times 3$, 
which yields an output of size~$393\times 393\times 32$.
There are four more convolutional layers which have output sizes of~$194 \times 194\times 64$, 
$95 \times 95\times 128$, $45 \times 45\times 256$, and~$20 \times 20\times 512$, respectively.
Each convolutional layer is followed by a max pooling layer and the final output of all these
layers is~$10\times 10 \times 512$. This is then flattened and forwarded to three dense layers for classification.
The CNN architecture that we use for the BODMAS dataset is illustrated in 
Figure~\ref{fig:CNN_both}(b).    
\end{itemize}

\section{Experiments and Results}\label{chap:RESULTS}

This section is split into two main parts, each describing the results for one of the two datasets that we consider. 
We summarize the overall results at the end of this section and we discuss our main findings. 

\subsection{CIC-MalMem-2022 Results}\label{sect:resCIC}

As discussed above, for the CIC-MalMem-2022 dataset, each sample consists of~55 features.
We first consider feature analysis to understand the relative importance of features in the dataset.

For feature selection, we use SelectKBest from \texttt{scikit-learn}, with the~$\chi^2$ option. 
SelectKBest determines the top~$K$ features that have the highest~$\chi^2$ values with respect to the target variable. 
The~$\chi^2$ test measures the dependence between stochastic variables, making this method suitable for 
determining the statistical significance of features. The top~10 features of highest importance in the CIC-MalMem-2022 dataset
are listed in Table~\ref{table:Top 10 features_CIC}; a complete list of all~55 features is
given in Table~\ref{tab:longtable} in Appendix~\ref{appendix_a}.

\begin{table}[!htb]
    \centering \def\z{\phantom{0}}
    \caption{Top~10 features for CIC-MalMem-2022 dataset}\label{table:Top 10 features_CIC}
    \begin{tabular}{c|c}
    \toprule
     Rank & Feature name  \\ \midrule
    \z1 & \texttt{malfind.commitCharge}  \\ 
    \z2 & \texttt{handles.nhandles}   \\
    \z3 & \texttt{handles.nevent}   \\
    \z4 & \texttt{handles.nsection}   \\
    \z5 & \texttt{handles.nthread}  \\
    \z6 & \texttt{dlllist.ndlls}   \\
    \z7 & \texttt{handles.nfile}  \\
    \z8 & \texttt{handles.nkey}  \\
    \z9 & \texttt{handles.nmutant}   \\
    10 & \texttt{handles.nsemaphore}  \\ \bottomrule
\end{tabular}
\end{table}

\subsubsection{Classic Learning Techniques}

In this section, we consider experiments involving Random Forest, SVC, and MLP classifiers.
The following experiments were conducted using the entire set of~55 features.
We also experimented with the reduced set of~10 features in Table~\ref{table:Top 10 features_CIC},
but the results are similar. 

\begin{figure}[!htb]
    \centering 
    \begin{tabular}{cc}
    %\begin{tikzpicture}[scale=0.8,every node/.style={scale=0.8}]
\begin{tikzpicture}[scale=0.3]
    \begin{axis}[%colorbar/width=2.5mm,
        width=15cm,
        height=15cm,
%        xlabel={\LARGE Predicted class},
%        ylabel={\LARGE Actual class},
%        colormap={blackwhite}{gray(0cm)=(1); gray(1cm)=(0.5)},
%	colormap={bluewhite}{color=(white) color=(blue)},
%	colormap={bluewhite}{color=(white) rgb255=(0,191,255)},
	colormap={bluewhite}{color=(white) rgb255=(100,149,237)},
        xticklabels={
Benign,
Spyware,
Ransomware,
Trojan
        },
        xtick={0,...,3},
        xtick style={draw=none},
	xticklabel style={anchor=east,rotate=60,yshift=-5pt,font=\large,scale=2.0},
        yticklabels={
Benign,
Spyware,
Ransomware,
Trojan
        },
        ytick={0,...,3},
        ytick style={draw=none},
        enlargelimits=false,
        yticklabel style={font=\large,scale=2.0},
        colorbar,
        colorbar style={
%     	  	width=0.05*\pgfkeysvalueof{/pgfplots/parent axis width},%%% added this
%     	  	height=0.5*\pgfkeysvalueof{/pgfplots/parent axis height},
%		plot graphics/node/.style={scale=1.33,anchor=south west,inner sep=0pt,}, %%% scale colorbar fill %%%
%            ytick={0.00,0.20,0.40,0.60,0.80,1.00},
%            yticklabels={0.00,0.20,0.40,0.60,0.80,1.00},
            ytick={0,200,400,600,800,1000,1200},
            yticklabels={0,200,400,600,1000,1200},
            yticklabel={\pgfmathprintnumber\tick},
            yticklabel style={%font=\footnotesize,
            		scale=2.0,
            		/pgf/number format/fixed,
			/pgf/number format/fixed zerofill,
			/pgf/number format/1000 sep={},
%			/pgf/number format/precision=2}
			/pgf/number format/precision=0}
        },
        point meta min=0,
        point meta max=1250,
%        point meta min=0.0,
%        point meta max=1.0,
%        nodes near coords={\pgfmathprintnumber\pgfplotspointmeta\,\%},
        nodes near coords={\pgfmathprintnumber\pgfplotspointmeta},
        % ---------------------------------------------------------------------
        % show `nodes near coords' but adapt the style so that values
        % above a threshold get another style
        % (adapted from <http://tex.stackexchange.com/a/141006/95441>)
        % #1: the THRESHOLD after which we switch to a special display.
        nodes near coords black white/.style={
            % define the style of the nodes with "small" values
            small value/.style={
                yshift=-14pt,
%                text=white,
                text=black,
                /pgf/number format/fixed,
                /pgf/number format/1000 sep={},
                /pgf/number format/precision=0,
%                /pgf/number format/precision=4,
                /pgf/number format/zerofill=true,
                scale=2.75,
%                /pgf/number format/precision=0
            },
            % define the style of the nodes with "large" values
            large value/.style={
                yshift=-14pt,
%                text=black,
                text=white,
                /pgf/number format/fixed,
                /pgf/number format/1000 sep={},
                /pgf/number format/precision=0,
%                /pgf/number format/precision=4,
                /pgf/number format/zerofill=true,
                scale=2.75,
%                /pgf/number format/precision=0
            },
            every node near coord/.style={
                check for zero/.code={
                    \pgfmathfloatifflags{\pgfplotspointmeta}{0}{
                        % If meta=0, make the node a coordinate
                        % (which doesn't have text)
                        \pgfkeys{/tikz/coordinate}
                    }{
                        \begingroup
                        % this group is merely to switch to FPU locally.
                        % Might be unnecessary, but who knows.
                        \pgfkeys{/pgf/fpu}
                        \pgfmathparse{\pgfplotspointmeta<#1}
                        \global\let\result=\pgfmathresult
                        \endgroup
                        %
                        % simplifies debugging:
                        %\show\result
                        %
                        \pgfmathfloatcreate{1}{1.0}{0}
                        \let\ONE=\pgfmathresult
                        \ifx\result\ONE
                            % AH: our condition 'y < #1' is met.
                            \pgfkeysalso{/pgfplots/small value}
                        \else
                            % ok, proceed as usual.
                            \pgfkeysalso{/pgfplots/large value}
                        \fi
                    }
                },
                check for zero,
            },
        },
        % asign a value to the new style which is the threshold at which
        % the two style `small value' or `large value' are used
        nodes near coords black white=625,
%        nodes near coords black white=0.5,
        % -----------------------------------------------------------------
    ]
        \addplot[
            matrix plot,
            mesh/cols=4,
            point meta=explicit,draw=gray
        ] table [meta=C] {
            x y C
0 0 1250
1 0 0
2 0 0
3 0 0
0 1 0
1 1 843
2 1 160
3 1 223
0 2 0
1 2 151
2 2 898
3 2 141
0 3 0
1 3 205
2 3 90
3 3 839
         };
    \end{axis}
%\draw[black,thick] (2.625,5.8) circle(0.5);
%\draw[red,dashed,thick] (3.675,5.8) circle(0.5);
\end{tikzpicture}
    &
    %\begin{tikzpicture}[scale=0.8,every node/.style={scale=0.8}]
\begin{tikzpicture}[scale=0.3]
    \begin{axis}[%colorbar/width=2.5mm,
        width=15cm,
        height=15cm,
%        xlabel={\LARGE Predicted class},
%        ylabel={\LARGE Actual class},
%        colormap={blackwhite}{gray(0cm)=(1); gray(1cm)=(0.5)},
%	colormap={bluewhite}{color=(white) color=(blue)},
%	colormap={bluewhite}{color=(white) rgb255=(0,191,255)},
	colormap={bluewhite}{color=(white) rgb255=(100,149,237)},
        xticklabels={
Benign,
Spyware,
Ransomware,
Trojan
        },
        xtick={0,...,3},
        xtick style={draw=none},
	xticklabel style={anchor=east,rotate=60,yshift=-5pt,font=\large,scale=2.0},
        yticklabels={
Benign,
Spyware,
Ransomware,
Trojan
        },
        ytick={0,...,3},
        ytick style={draw=none},
        enlargelimits=false,
        yticklabel style={font=\large,scale=2.0},
        colorbar,
        colorbar style={
%     	  	width=0.05*\pgfkeysvalueof{/pgfplots/parent axis width},%%% added this
%     	  	height=0.5*\pgfkeysvalueof{/pgfplots/parent axis height},
%		plot graphics/node/.style={scale=1.33,anchor=south west,inner sep=0pt,}, %%% scale colorbar fill %%%
%            ytick={0.00,0.20,0.40,0.60,0.80,1.00},
%            yticklabels={0.00,0.20,0.40,0.60,0.80,1.00},
            ytick={0,200,400,600,800,1000,1200},
            yticklabels={0,200,400,600,1000,1200},
            yticklabel={\pgfmathprintnumber\tick},
            yticklabel style={%font=\footnotesize,
            		scale=2.0,
            		/pgf/number format/fixed,
			/pgf/number format/fixed zerofill,
			/pgf/number format/1000 sep={},
%			/pgf/number format/precision=2}
			/pgf/number format/precision=0}
        },
        point meta min=0,
        point meta max=1250,
%        point meta min=0.0,
%        point meta max=1.0,
%        nodes near coords={\pgfmathprintnumber\pgfplotspointmeta\,\%},
        nodes near coords={\pgfmathprintnumber\pgfplotspointmeta},
        % ---------------------------------------------------------------------
        % show `nodes near coords' but adapt the style so that values
        % above a threshold get another style
        % (adapted from <http://tex.stackexchange.com/a/141006/95441>)
        % #1: the THRESHOLD after which we switch to a special display.
        nodes near coords black white/.style={
            % define the style of the nodes with "small" values
            small value/.style={
                yshift=-14pt,
%                text=white,
                text=black,
                /pgf/number format/fixed,
                /pgf/number format/1000 sep={},
                /pgf/number format/precision=0,
%                /pgf/number format/precision=4,
                /pgf/number format/zerofill=true,
                scale=2.75,
%                /pgf/number format/precision=0
            },
            % define the style of the nodes with "large" values
            large value/.style={
                yshift=-14pt,
%                text=black,
                text=white,
                /pgf/number format/fixed,
                /pgf/number format/1000 sep={},
                /pgf/number format/precision=0,
%                /pgf/number format/precision=4,
                /pgf/number format/zerofill=true,
                scale=2.75,
%                /pgf/number format/precision=0
            },
            every node near coord/.style={
                check for zero/.code={
                    \pgfmathfloatifflags{\pgfplotspointmeta}{0}{
                        % If meta=0, make the node a coordinate
                        % (which doesn't have text)
                        \pgfkeys{/tikz/coordinate}
                    }{
                        \begingroup
                        % this group is merely to switch to FPU locally.
                        % Might be unnecessary, but who knows.
                        \pgfkeys{/pgf/fpu}
                        \pgfmathparse{\pgfplotspointmeta<#1}
                        \global\let\result=\pgfmathresult
                        \endgroup
                        %
                        % simplifies debugging:
                        %\show\result
                        %
                        \pgfmathfloatcreate{1}{1.0}{0}
                        \let\ONE=\pgfmathresult
                        \ifx\result\ONE
                            % AH: our condition 'y < #1' is met.
                            \pgfkeysalso{/pgfplots/small value}
                        \else
                            % ok, proceed as usual.
                            \pgfkeysalso{/pgfplots/large value}
                        \fi
                    }
                },
                check for zero,
            },
        },
        % asign a value to the new style which is the threshold at which
        % the two style `small value' or `large value' are used
        nodes near coords black white=625,
%        nodes near coords black white=0.5,
        % -----------------------------------------------------------------
    ]
        \addplot[
            matrix plot,
            mesh/cols=4,
            point meta=explicit,draw=gray
        ] table [meta=C] {
            x y C
0 0 1220
1 0 1
2 0 8
3 0 21
0 1 3
1 1 28
2 1 192
3 1 1003
0 2 8
1 2 10
2 2 239
3 2 933
0 3 4
1 3 1
2 3 51
3 3 1078
         };
    \end{axis}
%\draw[black,thick] (2.625,5.8) circle(0.5);
%\draw[red,dashed,thick] (3.675,5.8) circle(0.5);
\end{tikzpicture}
    \\
    (a) Random Forest 
    &
    (b) SVC
    \\ \\[-1.0ex]
    \multicolumn{2}{c}{%\begin{tikzpicture}[scale=0.8,every node/.style={scale=0.8}]
\begin{tikzpicture}[scale=0.3]
    \begin{axis}[%colorbar/width=2.5mm,
        width=15cm,
        height=15cm,
%        xlabel={\LARGE Predicted class},
%        ylabel={\LARGE Actual class},
%        colormap={blackwhite}{gray(0cm)=(1); gray(1cm)=(0.5)},
%	colormap={bluewhite}{color=(white) color=(blue)},
%	colormap={bluewhite}{color=(white) rgb255=(0,191,255)},
	colormap={bluewhite}{color=(white) rgb255=(100,149,237)},
        xticklabels={
Benign,
Spyware,
Ransomware,
Trojan
        },
        xtick={0,...,3},
        xtick style={draw=none},
	xticklabel style={anchor=east,rotate=60,yshift=-5pt,font=\large,scale=2.0},
        yticklabels={
Benign,
Spyware,
Ransomware,
Trojan
        },
        ytick={0,...,3},
        ytick style={draw=none},
        enlargelimits=false,
        yticklabel style={font=\large,scale=2.0},
        colorbar,
        colorbar style={
%     	  	width=0.05*\pgfkeysvalueof{/pgfplots/parent axis width},%%% added this
%     	  	height=0.5*\pgfkeysvalueof{/pgfplots/parent axis height},
%		plot graphics/node/.style={scale=1.33,anchor=south west,inner sep=0pt,}, %%% scale colorbar fill %%%
%            ytick={0.00,0.20,0.40,0.60,0.80,1.00},
%            yticklabels={0.00,0.20,0.40,0.60,0.80,1.00},
            ytick={0,200,400,600,800,1000,1200},
            yticklabels={0,200,400,600,1000,1200},
            yticklabel={\pgfmathprintnumber\tick},
            yticklabel style={%font=\footnotesize,
            		scale=2.5,
            		/pgf/number format/fixed,
			/pgf/number format/fixed zerofill,
			/pgf/number format/1000 sep={},
%			/pgf/number format/precision=2}
			/pgf/number format/precision=0}
        },
        point meta min=0,
        point meta max=1250,
%        point meta min=0.0,
%        point meta max=1.0,
%        nodes near coords={\pgfmathprintnumber\pgfplotspointmeta\,\%},
        nodes near coords={\pgfmathprintnumber\pgfplotspointmeta},
        % ---------------------------------------------------------------------
        % show `nodes near coords' but adapt the style so that values
        % above a threshold get another style
        % (adapted from <http://tex.stackexchange.com/a/141006/95441>)
        % #1: the THRESHOLD after which we switch to a special display.
        nodes near coords black white/.style={
            % define the style of the nodes with "small" values
            small value/.style={
                yshift=-14pt,
%                text=white,
                text=black,
                /pgf/number format/fixed,
                /pgf/number format/1000 sep={},
                /pgf/number format/precision=0,
%                /pgf/number format/precision=4,
                /pgf/number format/zerofill=true,
                scale=2.75,
%                /pgf/number format/precision=0
            },
            % define the style of the nodes with "large" values
            large value/.style={
                yshift=-14pt,
%                text=black,
                text=white,
                /pgf/number format/fixed,
                /pgf/number format/1000 sep={},
                /pgf/number format/precision=0,
%                /pgf/number format/precision=4,
                /pgf/number format/zerofill=true,
                scale=2.75,
%                /pgf/number format/precision=0
            },
            every node near coord/.style={
                check for zero/.code={
                    \pgfmathfloatifflags{\pgfplotspointmeta}{0}{
                        % If meta=0, make the node a coordinate
                        % (which doesn't have text)
                        \pgfkeys{/tikz/coordinate}
                    }{
                        \begingroup
                        % this group is merely to switch to FPU locally.
                        % Might be unnecessary, but who knows.
                        \pgfkeys{/pgf/fpu}
                        \pgfmathparse{\pgfplotspointmeta<#1}
                        \global\let\result=\pgfmathresult
                        \endgroup
                        %
                        % simplifies debugging:
                        %\show\result
                        %
                        \pgfmathfloatcreate{1}{1.0}{0}
                        \let\ONE=\pgfmathresult
                        \ifx\result\ONE
                            % AH: our condition 'y < #1' is met.
                            \pgfkeysalso{/pgfplots/small value}
                        \else
                            % ok, proceed as usual.
                            \pgfkeysalso{/pgfplots/large value}
                        \fi
                    }
                },
                check for zero,
            },
        },
        % asign a value to the new style which is the threshold at which
        % the two style `small value' or `large value' are used
        nodes near coords black white=625,
%        nodes near coords black white=0.5,
        % -----------------------------------------------------------------
    ]
        \addplot[
            matrix plot,
            mesh/cols=4,
            point meta=explicit,draw=gray
        ] table [meta=C] {
            x y C
0 0 1242
1 0 8
2 0 0
3 0 0
0 1 2
1 1 589
2 1 21
3 1 614
0 2 21
1 2 579
2 2 26
3 2 564
0 3 2
1 3 387
2 3 14
3 3 731
         };
    \end{axis}
%\draw[black,thick] (2.625,5.8) circle(0.5);
%\draw[red,dashed,thick] (3.675,5.8) circle(0.5);
\end{tikzpicture}}
    \\
    \multicolumn{2}{c}{(c) MLP}
    \end{tabular}
    \caption{Confusion matrices for classic techniques (CIC-MalMem-2022)}
    \label{fig:Confusion Matrix for Random Forest}
\end{figure}
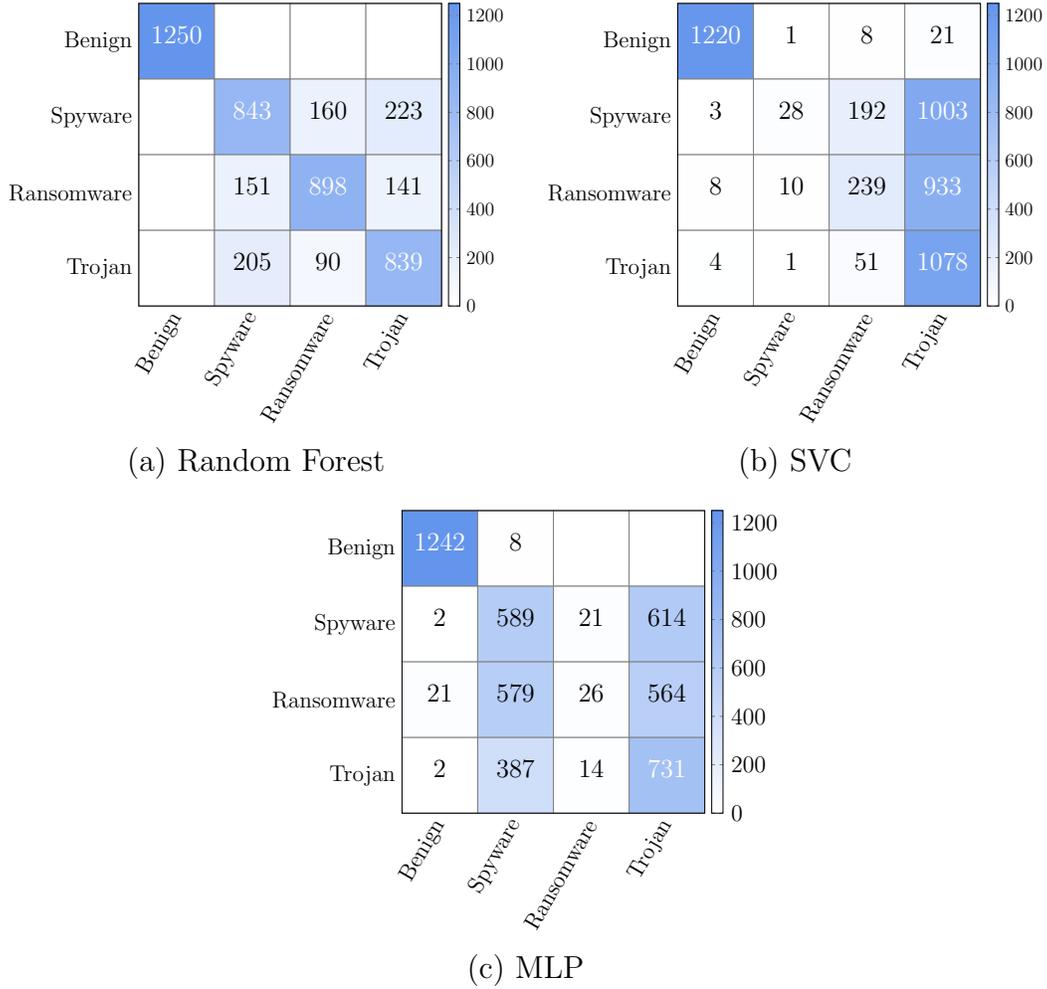

When trained on the entire set of~55 features with~1000 estimators, Random Forest
achieves an accuracy of~0.7979. From the confusion matrix in Figure~\ref{fig:Confusion Matrix for Random Forest}(a),
we observe that all the benign examples were classified correctly; 
however the classifier had considerable difficulty distinguishing between the three 
malware categories. The F1-score for this classifier is~0.7980

When trained on the set of~55 features, SVC achieves an accuracy of~0.5343. 
The confusion matrix in Figure~\ref{fig:Confusion Matrix for Random Forest}(b)
further emphasizes the poor performance of this model. 
The F1-score achieved using this method is~0.4597 which is also much worse than the Random Forest.

When trained on the full~55 features, our MLP classifier achieves~0.5391 accuracy. 
The confusion matrix in Figure~\ref{fig:Confusion Matrix for Random Forest}(c) 
shows that this MLP model tends to make different types of mistakes than the SVC.
The F1-scores achieved by our MLP model is~0.4886, 
which is similar to that of the SVC. 

Table~\ref{table:Accuraciesl} summarizes the accuracies and F1-scores
for the Random Forest, SVC, and MLP classifiers, both with and without feature reduction.
We observe that feature selection has little effect on the performance of these models.

\begin{table}[!htb]
    \centering
    \caption{Accuracies and F1-scores (CIC-MalMem-2022 dataset)}
    \label{table:Accuraciesl}
\begin{tabular}{cc|ccc}
    \toprule
  Measure & Features & Random Forest & SVC  & MLP  \\ \midrule
\multirow{2}{*}{Accuracy} & 55 & 0.7979 & \phantom{0}0.5343\phantom{0}  & \phantom{0}0.5391\phantom{0} \\
    & 10 & 0.7689 & 0.5395 & 0.5483 \\ \midrule
\multirow{2}{*}{F1-score} & 55 & 0.7980 & 0.4597 & 0.4886 \\
    & 10 & 0.7690 & 0.4615 & 0.4710 \\ 
    \bottomrule
\end{tabular}
\end{table}

\subsubsection{QR Code Experiments}

\begin{figure}[!htb]
    \centering
    \begin{tabular}{ccc}
    \includegraphics[width=35mm]{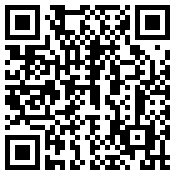}
    & &
    \includegraphics[width=35mm]{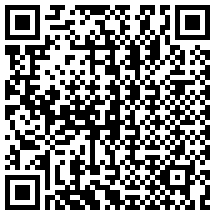}
    \\
    (a) Benign
    & &
    (b) Ransomware
    \\ \\[-1.5ex]
    \includegraphics[width=35mm]{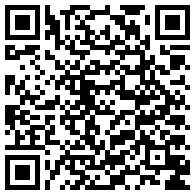}
    & &
    \includegraphics[width=35mm]{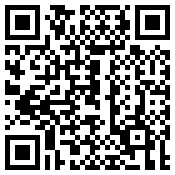}
    \\
    (c) Spyware
    & &
    (d) Trojan
    \end{tabular}
    \caption{Examples of QR code representations (CIC-MalMem-2022)}
    \label{fig: MALMEM_BENIGN}
\end{figure}

To create our QR representations for the CIC-MalMem-2022 dataset, the top~10 features were used. 
Figures~\ref{fig: MALMEM_BENIGN}(a) through~(d) give examples of QR representations of benign, ransomware,
spyware, and Trojan, respectively. 

The parameters used to generate the QR codes for this dataset are given in 
Table~\ref{table:QR Parameters for CIC-MalMem}.
The generated images were of size~$175 \times 175$ pixels each. 
These were then resized to images of size~$128 \times 128$ before being used as input to the CNN. 

\begin{table}[!htb]
    \centering
    \caption{QR parameters for CIC-MalMem-2022}
    \label{table:QR Parameters for CIC-MalMem}
    \begin{tabular}{c|c}
    \toprule
    Parameter & Value \\ \midrule
    \texttt{version} & 1 \\
    \texttt{error\_correction} & \texttt{ERROR\_CORRECT\_L}\\
    \texttt{box\_size} & 5 \\
    \texttt{border} & 1 \\ \bottomrule
\end{tabular}
\end{table}

\begin{figure}[!htb]
    \centering
    \includegraphics[width=120mm]{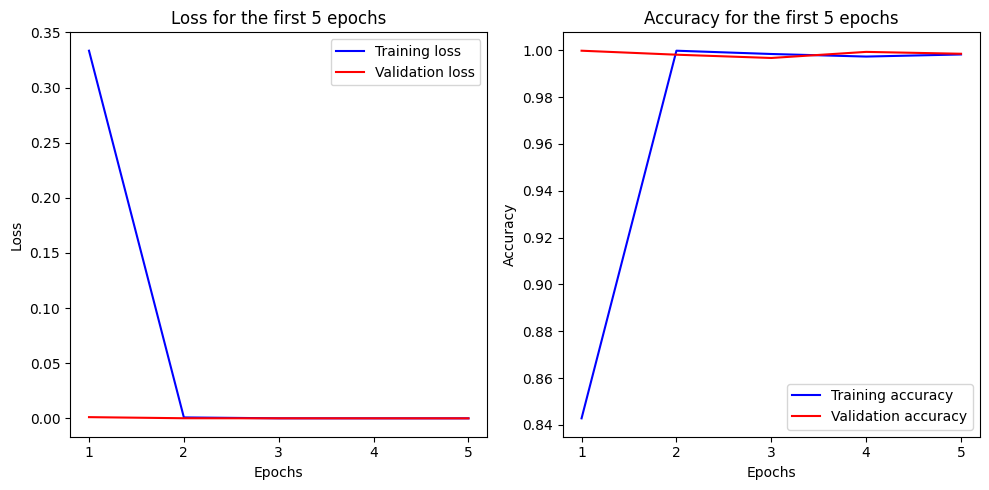}
    \caption{QR-CNN accuracy and loss graphs for CIC-MalMem-2022}
    \label{fig: MALMEM_GRAPH}
\end{figure}

% Here it says 24k, but previously it was 20k...
We split the~24,000 samples~70:15:15 for train:validation:test. 
The train and validation split was used to train on the 
CNN model described in Section~\ref{chap:METHOD}.
The loss and accuracy graphs are shown in Figure~\ref{fig: MALMEM_GRAPH}.
These graphs show that the model converged after two epochs, with
no signs of overfitting. The test accuracy achieved for the CNN on the QR 
image representation was~0.9998 
for this multiclass classification problem.

\subsubsection{Aztec Code Experiments}

\begin{figure}[!htb]
    \centering
    \begin{tabular}{ccc}
    \includegraphics[width=35mm]{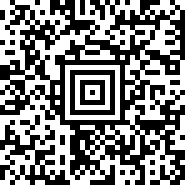}
    & &
    \includegraphics[width=35mm]{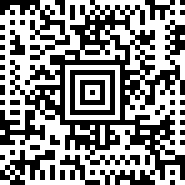}
    \\
    (a) Benign
    & &
    (b) Ransomware
    \\ \\[-1.5ex]
    \includegraphics[width=35mm]{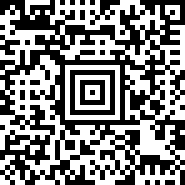}
    & &
    \includegraphics[width=35mm]{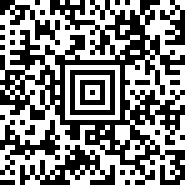}
    \\
    (c) Spyware
    & &
    (d) Trojan
    \end{tabular}
    \caption{Examples of Aztec code representations (CIC-MalMem-2022)}
    \label{fig: MALMEM_BENIGN_Aztec}
\end{figure}

To create the Aztec code representation, as for the QR codes,
we use the top~10 features. Figures~\ref{fig: MALMEM_BENIGN_Aztec}(a) through~(d) 
give typical examples of Aztec code representations for the 
benign, ransomware,
spyware, and Trojan classes, respectively. 

The only parameter that we employ when generating our Aztec code
representations is~$\texttt{module\_size} = 5$.
The generated images are of size~$175 \times 175$ pixels, and these were then resized 
to images of size~$128 \times 128$ so as to be suitable as input to the CNN. 

\begin{figure}[!htb]
    \centering
    \includegraphics[width=120mm]{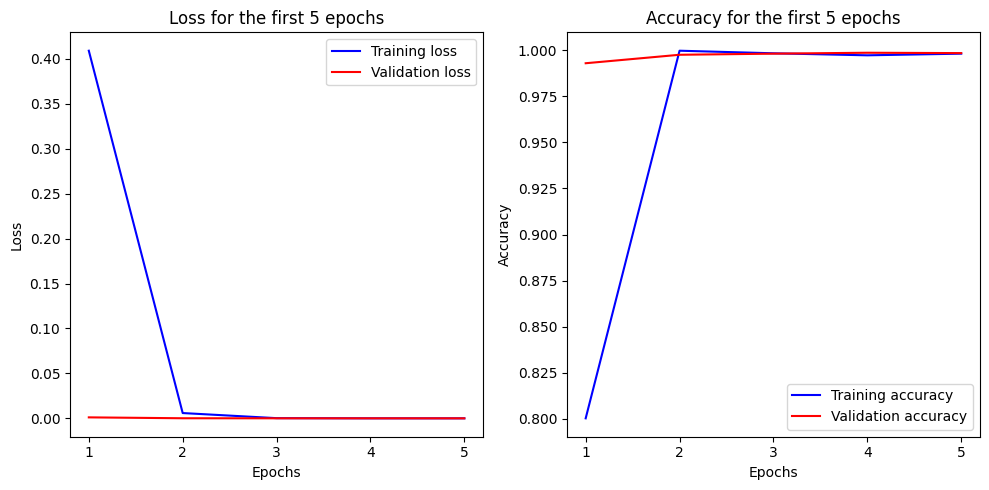}
    \caption{Aztec-CNN accuracy and loss graphs for CIC-MalMem-2022}
    \label{fig: MALMEM_AZTEC_GRAPH}
\end{figure}

We split the 
%24,000 selected 
samples~70:15:15 for train:validation:test. 
The train and validation split was used to train the CNN architecture
described in Section~\ref{chap:METHOD}.
The loss and accuracy graphs for this model
are shown in Figure~\ref{fig: MALMEM_AZTEC_GRAPH}.
As with the QR-CNN results in Figure~\ref{fig: MALMEM_GRAPH},
the graphs in Figure~\ref{fig: MALMEM_AZTEC_GRAPH}
show that the model converges after
two epochs, with no indication of overfitting.
 
The test accuracy achieved for the CNN on the QR image representation 
was~0.9986 for this multiclass classification problem. 
While this is marginally less than the accuracy achieved using the QR code representation,
both represent nearly perfect classification.

\subsection{BODMAS Results}\label{sect:resBOD}

In this section, we conduct analogous experiments as the previous section, but
based on the BODMAS dataset, rather than the CIC-MalMem-2022 dataset. 
We first discuss feature selection before turning our attention to our experimental results.

\subsubsection{Feature Selection}

Recall that each BODMAS sample consists of a~2384 dimensional vector,
which was extracted using the LIEF project. For the classic ML models, we experiment with the
top~50 and the top~150 features. Similarly, our CNN models are trained on QR and Aztec images
derived from the same top~50 and top~150 features.

\begin{figure}[!htb]
    \centering
    \begin{tikzpicture}[scale=0.85, every node/.style={scale=1.0}]
\pgfkeys{/pgf/number format/.cd,1000 sep={}}
\begin{axis}[%bar shift=0pt,
        width  = 0.925*\textwidth,
        height = 7.0cm,
        ymin=0, ymax=60,
%        xmin=0.0005, xmax=0.040,
%        ytick={0.0, 0.1, 0.2, 0.3, 0.4, 0.5, 0.6, 0.7, 0.8, 0.9, 1.0},
%        ytick={0, 5000, 10000, 15000, 20000, 25000, 30000},
%        ytick scale label code/.code={},
        major x tick style = transparent,
        ybar=5*\pgflinewidth,
        bar width=12pt,
%        ymajorgrids = true,
%        xlabel = {Learning technique},
        ylabel = {Frequency},
        symbolic x coords={0.0010, 0.0028, 0.0046, 0.0064, 0.0083, 0.0101, 0.0119, 0.0137, 0.0156, 
        		0.0174, 0.0192, 0.0210, 0.0228, 0.0247, 0.0265, 0.0283, 0.0301, 0.0320, 0.0338, 0.0356},
        xticklabels={0.0010, 0.0028, 0.0046, 0.0064, 0.0083, 0.0101, 0.0119, 0.0137, 0.0156, 
        		0.0174, 0.0192, 0.0210, 0.0228, 0.0247, 0.0265, 0.0283, 0.0301, 0.0320, 0.0338, 0.0356},
	y tick label style={
    		/pgf/number format/.cd,
   		fixed,
   		fixed zerofill,
%		sep=,
    		precision=0
	},
%	yticklabel pos=right,
        xtick = data,
        x tick label style={
              rotate=60,
		anchor=north east,
    		/pgf/number format/.cd,
   		fixed,
   		fixed zerofill,
%		sep=,
    		precision=3
		},
%		font=\small},
%        scaled y ticks = false,
	%%%%% numbers on bars and rotated
%        nodes near coords,
%        every node near coord/.append style={rotate=90, 
%        								   anchor=west,
%								   %font=\footnotesize,
%								   /pgf/number format/.cd,
%								   fixed,
%								   fixed zerofill,
%%								   sep=,
%								   precision=4},
        %%%%%
%        enlarge x limits=0.03,
%        enlarge x limits=0.06,
        enlarge x limits=0.04,
        legend cell align=left,
        legend pos=north east,
%        every axis plot/.append style={
%        		bar shift=0pt,
%		fill
%        },
%        legend style={
%%                at={(1,1.05)},
%%                anchor=south east,
%%	        nodes={rotate=90},%%%%% rotate text in legend
%%                at={(0.125,0)},
%%                at={(0.125,0)},
%%                at={(0.8775,0)},
%                at={(0.82,0.56)},
%                anchor=south,
%                column sep=1ex
%        },
]
\addplot [fill=blue,opacity=1.00]
coordinates {
(0.0010, 57)
(0.0028, 24)
(0.0046, 12) 
(0.0064, 10)
(0.0083, 4)
(0.0101, 9)
(0.0119, 5)
(0.0137, 5)
(0.0156, 5)
(0.0174, 2)
(0.0192, 1)
(0.0210, 5)
(0.0228, 2)
(0.0247, 6)
(0.0265, 2)
(0.0283, 0)
(0.0301, 0)
(0.0320, 0)
(0.0338, 0)
(0.0356, 1)
};
\end{axis}
\end{tikzpicture}
    \caption{Feature importance distribution in BODMAS}\label{fig:Feature_Importance_Bodmas}
\end{figure}
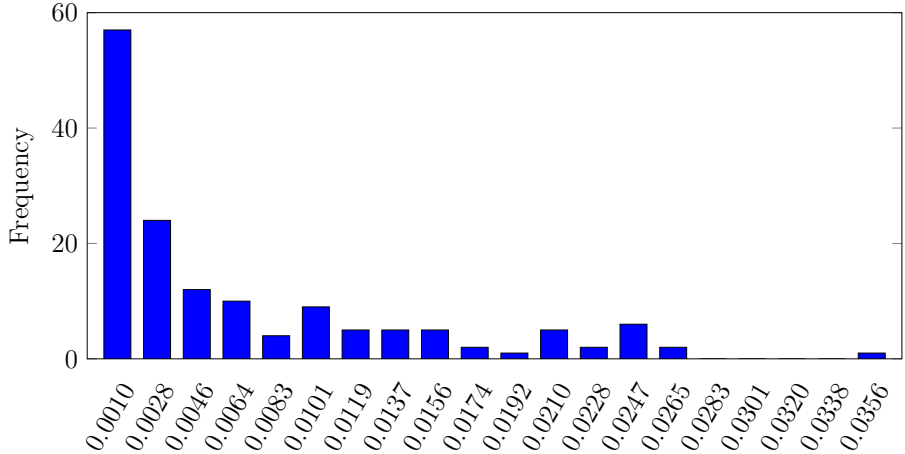

Figure~\ref{fig:Feature_Importance_Bodmas} shows the distribution of the 
top~150 features among the~2384 BODMAS features. This figure highlights the fact that 
most of the features are of little---if any---relevance for classification. 

%\begin{figure}[!htb]
%    \centering
%    \includegraphics[width=85mm]{./images/Feature_Importance_Bodmas.png}
%    \caption{Feature importance in BODMAS}
%    \label{fig:Feature_Importance_Bodmas}
%\end{figure}

The feature selection for the BODMAS dataset was done by selecting the top~50 features using 
\texttt{SelectKBest} with ANOVA as the a statistical technique.
The~10 most significant features for the BODMAS dataset are shown in 
Table~\ref{table:Top 10 features_BODMAS}. 

\begin{table}[!htb]
    \centering \def\z{\phantom{0}}
    \caption{Top~10 features for BODMAS dataset}\label{table:Top 10 features_BODMAS}
    \begin{tabular}{c|cc}
    \toprule
     Rank & Feature number & Importance  \\ \midrule
    \z1 & \z584 & 0.036483\\
    \z2 & \z473 & 0.026751\\
    \z3 & 1283 & 0.025871\\
    \z4 & \z137 & 0.024794\\
    \z5 & \z\z44 & 0.024691\\
    \z6 & \z506 & 0.024518 \\
    \z7 & \z\z62 & 0.024090 \\
    \z8 & \z\z38 & 0.024042 \\
    \z9 & \z499 & 0.023880  \\
    10 & \z\z27 & 0.022573 \\ \bottomrule
    \end{tabular}
\end{table}

\subsubsection{Classic Learning Techniques}

When trained on the set of~50 features with~1000 estimators, the Random Forest achieves an accuracy of~0.946. 
From the confusion matrix in Figure~\ref{fig:CM_RF}(a), we observe that almost all the benign examples are classified 
correctly, while the classifier has more difficulty with the three malware classes.

\begin{figure}[!htb]
    \centering 
    \begin{tabular}{cc}
    %\begin{tikzpicture}[scale=0.8,every node/.style={scale=0.8}]
\begin{tikzpicture}[scale=0.3]
    \begin{axis}[%colorbar/width=2.5mm,
        width=15cm,
        height=15cm,
%        xlabel={\LARGE Predicted class},
%        ylabel={\LARGE Actual class},
%        colormap={blackwhite}{gray(0cm)=(1); gray(1cm)=(0.5)},
%	colormap={bluewhite}{color=(white) color=(blue)},
%	colormap={bluewhite}{color=(white) rgb255=(0,191,255)},
	colormap={bluewhite}{color=(white) rgb255=(100,149,237)},
        xticklabels={
Benign,
\texttt{Upatre},
\texttt{Wacatac},
\texttt{Sfone}
        },
        xtick={0,...,3},
        xtick style={draw=none},
	xticklabel style={anchor=east,rotate=60,yshift=-5pt,font=\large,scale=2.0},
        yticklabels={
Benign,
\texttt{Upatre},
\texttt{Wacatac},
\texttt{Sfone}
        },
        ytick={0,...,3},
        ytick style={draw=none},
        enlargelimits=false,
        yticklabel style={font=\large,scale=2.0},
        colorbar,
        colorbar style={
%     	  	width=0.05*\pgfkeysvalueof{/pgfplots/parent axis width},%%% added this
%     	  	height=0.5*\pgfkeysvalueof{/pgfplots/parent axis height},
%		plot graphics/node/.style={scale=1.33,anchor=south west,inner sep=0pt,}, %%% scale colorbar fill %%%
%            ytick={0.00,0.20,0.40,0.60,0.80,1.00},
%            yticklabels={0.00,0.20,0.40,0.60,0.80,1.00},
            ytick={0,200,400,600,800,1000},
            yticklabels={0,200,400,600,1000},
            yticklabel={\pgfmathprintnumber\tick},
            yticklabel style={%font=\footnotesize,
            		scale=2.0,
            		/pgf/number format/fixed,
			/pgf/number format/fixed zerofill,
			/pgf/number format/1000 sep={},
%			/pgf/number format/precision=2}
			/pgf/number format/precision=0}
        },
        point meta min=0,
        point meta max=1000,
%        point meta min=0.0,
%        point meta max=1.0,
%        nodes near coords={\pgfmathprintnumber\pgfplotspointmeta\,\%},
        nodes near coords={\pgfmathprintnumber\pgfplotspointmeta},
        % ---------------------------------------------------------------------
        % show `nodes near coords' but adapt the style so that values
        % above a threshold get another style
        % (adapted from <http://tex.stackexchange.com/a/141006/95441>)
        % #1: the THRESHOLD after which we switch to a special display.
        nodes near coords black white/.style={
            % define the style of the nodes with "small" values
            small value/.style={
                yshift=-14pt,
%                text=white,
                text=black,
                /pgf/number format/fixed,
                /pgf/number format/1000 sep={},
                /pgf/number format/precision=0,
%                /pgf/number format/precision=4,
                /pgf/number format/zerofill=true,
                scale=2.75,
%                /pgf/number format/precision=0
            },
            % define the style of the nodes with "large" values
            large value/.style={
                yshift=-14pt,
%                text=black,
                text=white,
                /pgf/number format/fixed,
                /pgf/number format/1000 sep={},
                /pgf/number format/precision=0,
%                /pgf/number format/precision=4,
                /pgf/number format/zerofill=true,
                scale=2.75,
%                /pgf/number format/precision=0
            },
            every node near coord/.style={
                check for zero/.code={
                    \pgfmathfloatifflags{\pgfplotspointmeta}{0}{
                        % If meta=0, make the node a coordinate
                        % (which doesn't have text)
                        \pgfkeys{/tikz/coordinate}
                    }{
                        \begingroup
                        % this group is merely to switch to FPU locally.
                        % Might be unnecessary, but who knows.
                        \pgfkeys{/pgf/fpu}
                        \pgfmathparse{\pgfplotspointmeta<#1}
                        \global\let\result=\pgfmathresult
                        \endgroup
                        %
                        % simplifies debugging:
                        %\show\result
                        %
                        \pgfmathfloatcreate{1}{1.0}{0}
                        \let\ONE=\pgfmathresult
                        \ifx\result\ONE
                            % AH: our condition 'y < #1' is met.
                            \pgfkeysalso{/pgfplots/small value}
                        \else
                            % ok, proceed as usual.
                            \pgfkeysalso{/pgfplots/large value}
                        \fi
                    }
                },
                check for zero,
            },
        },
        % asign a value to the new style which is the threshold at which
        % the two style `small value' or `large value' are used
        nodes near coords black white=625,
%        nodes near coords black white=0.5,
        % -----------------------------------------------------------------
    ]
        \addplot[
            matrix plot,
            mesh/cols=4,
            point meta=explicit,draw=gray
        ] table [meta=C] {
            x y C
0 0 1015
1 0 0
2 0 5
3 0 6
0 1 0
1 1 926
2 1 0
3 1 22
0 2 16
1 2 0
2 2 731
3 2 5
0 3 22
1 3 86
2 3 10
3 3 827
         };
    \end{axis}
%\draw[black,thick] (2.625,5.8) circle(0.5);
%\draw[red,dashed,thick] (3.675,5.8) circle(0.5);
\end{tikzpicture}
    &
    %\begin{tikzpicture}[scale=0.8,every node/.style={scale=0.8}]
\begin{tikzpicture}[scale=0.3]
    \begin{axis}[%colorbar/width=2.5mm,
        width=15cm,
        height=15cm,
%        xlabel={\LARGE Predicted class},
%        ylabel={\LARGE Actual class},
%        colormap={blackwhite}{gray(0cm)=(1); gray(1cm)=(0.5)},
%	colormap={bluewhite}{color=(white) color=(blue)},
%	colormap={bluewhite}{color=(white) rgb255=(0,191,255)},
	colormap={bluewhite}{color=(white) rgb255=(100,149,237)},
        xticklabels={
Benign,
\texttt{Upatre},
\texttt{Wacatac},
\texttt{Sfone}
        },
        xtick={0,...,3},
        xtick style={draw=none},
	xticklabel style={anchor=east,rotate=60,yshift=-5pt,font=\large,scale=2.0},
        yticklabels={
Benign,
\texttt{Upatre},
\texttt{Wacatac},
\texttt{Sfone}
        },
        ytick={0,...,3},
        ytick style={draw=none},
        enlargelimits=false,
        yticklabel style={font=\large,scale=2.0},
        colorbar,
        colorbar style={
%     	  	width=0.05*\pgfkeysvalueof{/pgfplots/parent axis width},%%% added this
%     	  	height=0.5*\pgfkeysvalueof{/pgfplots/parent axis height},
%		plot graphics/node/.style={scale=1.33,anchor=south west,inner sep=0pt,}, %%% scale colorbar fill %%%
%            ytick={0.00,0.20,0.40,0.60,0.80,1.00},
%            yticklabels={0.00,0.20,0.40,0.60,0.80,1.00},
            ytick={0,200,400,600,800,1000},
            yticklabels={0,200,400,600,1000},
            yticklabel={\pgfmathprintnumber\tick},
            yticklabel style={%font=\footnotesize,
            		scale=2.0,
            		/pgf/number format/fixed,
			/pgf/number format/fixed zerofill,
			/pgf/number format/1000 sep={},
%			/pgf/number format/precision=2}
			/pgf/number format/precision=0}
        },
        point meta min=0,
        point meta max=1000,
%        point meta min=0.0,
%        point meta max=1.0,
%        nodes near coords={\pgfmathprintnumber\pgfplotspointmeta\,\%},
        nodes near coords={\pgfmathprintnumber\pgfplotspointmeta},
        % ---------------------------------------------------------------------
        % show `nodes near coords' but adapt the style so that values
        % above a threshold get another style
        % (adapted from <http://tex.stackexchange.com/a/141006/95441>)
        % #1: the THRESHOLD after which we switch to a special display.
        nodes near coords black white/.style={
            % define the style of the nodes with "small" values
            small value/.style={
                yshift=-14pt,
%                text=white,
                text=black,
                /pgf/number format/fixed,
                /pgf/number format/1000 sep={},
                /pgf/number format/precision=0,
%                /pgf/number format/precision=4,
                /pgf/number format/zerofill=true,
                scale=2.75,
%                /pgf/number format/precision=0
            },
            % define the style of the nodes with "large" values
            large value/.style={
                yshift=-14pt,
%                text=black,
                text=white,
                /pgf/number format/fixed,
                /pgf/number format/1000 sep={},
                /pgf/number format/precision=0,
%                /pgf/number format/precision=4,
                /pgf/number format/zerofill=true,
                scale=2.75,
%                /pgf/number format/precision=0
            },
            every node near coord/.style={
                check for zero/.code={
                    \pgfmathfloatifflags{\pgfplotspointmeta}{0}{
                        % If meta=0, make the node a coordinate
                        % (which doesn't have text)
                        \pgfkeys{/tikz/coordinate}
                    }{
                        \begingroup
                        % this group is merely to switch to FPU locally.
                        % Might be unnecessary, but who knows.
                        \pgfkeys{/pgf/fpu}
                        \pgfmathparse{\pgfplotspointmeta<#1}
                        \global\let\result=\pgfmathresult
                        \endgroup
                        %
                        % simplifies debugging:
                        %\show\result
                        %
                        \pgfmathfloatcreate{1}{1.0}{0}
                        \let\ONE=\pgfmathresult
                        \ifx\result\ONE
                            % AH: our condition 'y < #1' is met.
                            \pgfkeysalso{/pgfplots/small value}
                        \else
                            % ok, proceed as usual.
                            \pgfkeysalso{/pgfplots/large value}
                        \fi
                    }
                },
                check for zero,
            },
        },
        % asign a value to the new style which is the threshold at which
        % the two style `small value' or `large value' are used
        nodes near coords black white=625,
%        nodes near coords black white=0.5,
        % -----------------------------------------------------------------
    ]
        \addplot[
            matrix plot,
            mesh/cols=4,
            point meta=explicit,draw=gray
        ] table [meta=C] {
            x y C
0 0 989
1 0 0
2 0 19
3 0 18
0 1 9
1 1 933
2 1 0
3 1 6
0 2 14
1 2 0
2 2 728
3 2 10
0 3 59
1 3 111
2 3 51
3 3 724
         };
    \end{axis}
%\draw[black,thick] (2.625,5.8) circle(0.5);
%\draw[red,dashed,thick] (3.675,5.8) circle(0.5);
\end{tikzpicture}
    \\
    (a) Random Forest 
    &
    (b) SVC
    \\ \\[-1.0ex]
    \multicolumn{2}{c}{%\begin{tikzpicture}[scale=0.8,every node/.style={scale=0.8}]
\begin{tikzpicture}[scale=0.3]
    \begin{axis}[%colorbar/width=2.5mm,
        width=15cm,
        height=15cm,
%        xlabel={\LARGE Predicted class},
%        ylabel={\LARGE Actual class},
%        colormap={blackwhite}{gray(0cm)=(1); gray(1cm)=(0.5)},
%	colormap={bluewhite}{color=(white) color=(blue)},
%	colormap={bluewhite}{color=(white) rgb255=(0,191,255)},
	colormap={bluewhite}{color=(white) rgb255=(100,149,237)},
        xticklabels={
Benign,
\texttt{Upatre},
\texttt{Wacatac},
\texttt{Sfone}
        },
        xtick={0,...,3},
        xtick style={draw=none},
	xticklabel style={anchor=east,rotate=60,yshift=-5pt,font=\large,scale=2.0},
        yticklabels={
Benign,
\texttt{Upatre},
\texttt{Wacatac},
\texttt{Sfone}
        },
        ytick={0,...,3},
        ytick style={draw=none},
        enlargelimits=false,
        yticklabel style={font=\large,scale=2.0},
        colorbar,
        colorbar style={
%     	  	width=0.05*\pgfkeysvalueof{/pgfplots/parent axis width},%%% added this
%     	  	height=0.5*\pgfkeysvalueof{/pgfplots/parent axis height},
%		plot graphics/node/.style={scale=1.33,anchor=south west,inner sep=0pt,}, %%% scale colorbar fill %%%
%            ytick={0.00,0.20,0.40,0.60,0.80,1.00},
%            yticklabels={0.00,0.20,0.40,0.60,0.80,1.00},
            ytick={0,200,400,600,800,1000},
            yticklabels={0,200,400,600,1000},
            yticklabel={\pgfmathprintnumber\tick},
            yticklabel style={%font=\footnotesize,
            		scale=2.0,
            		/pgf/number format/fixed,
			/pgf/number format/fixed zerofill,
			/pgf/number format/1000 sep={},
%			/pgf/number format/precision=2}
			/pgf/number format/precision=0}
        },
        point meta min=0,
        point meta max=1000,
%        point meta min=0.0,
%        point meta max=1.0,
%        nodes near coords={\pgfmathprintnumber\pgfplotspointmeta\,\%},
        nodes near coords={\pgfmathprintnumber\pgfplotspointmeta},
        % ---------------------------------------------------------------------
        % show `nodes near coords' but adapt the style so that values
        % above a threshold get another style
        % (adapted from <http://tex.stackexchange.com/a/141006/95441>)
        % #1: the THRESHOLD after which we switch to a special display.
        nodes near coords black white/.style={
            % define the style of the nodes with "small" values
            small value/.style={
                yshift=-14pt,
%                text=white,
                text=black,
                /pgf/number format/fixed,
                /pgf/number format/1000 sep={},
                /pgf/number format/precision=0,
%                /pgf/number format/precision=4,
                /pgf/number format/zerofill=true,
                scale=2.75,
%                /pgf/number format/precision=0
            },
            % define the style of the nodes with "large" values
            large value/.style={
                yshift=-14pt,
%                text=black,
                text=white,
                /pgf/number format/fixed,
                /pgf/number format/1000 sep={},
                /pgf/number format/precision=0,
%                /pgf/number format/precision=4,
                /pgf/number format/zerofill=true,
                scale=2.75,
%                /pgf/number format/precision=0
            },
            every node near coord/.style={
                check for zero/.code={
                    \pgfmathfloatifflags{\pgfplotspointmeta}{0}{
                        % If meta=0, make the node a coordinate
                        % (which doesn't have text)
                        \pgfkeys{/tikz/coordinate}
                    }{
                        \begingroup
                        % this group is merely to switch to FPU locally.
                        % Might be unnecessary, but who knows.
                        \pgfkeys{/pgf/fpu}
                        \pgfmathparse{\pgfplotspointmeta<#1}
                        \global\let\result=\pgfmathresult
                        \endgroup
                        %
                        % simplifies debugging:
                        %\show\result
                        %
                        \pgfmathfloatcreate{1}{1.0}{0}
                        \let\ONE=\pgfmathresult
                        \ifx\result\ONE
                            % AH: our condition 'y < #1' is met.
                            \pgfkeysalso{/pgfplots/small value}
                        \else
                            % ok, proceed as usual.
                            \pgfkeysalso{/pgfplots/large value}
                        \fi
                    }
                },
                check for zero,
            },
        },
        % asign a value to the new style which is the threshold at which
        % the two style `small value' or `large value' are used
        nodes near coords black white=625,
%        nodes near coords black white=0.5,
        % -----------------------------------------------------------------
    ]
        \addplot[
            matrix plot,
            mesh/cols=4,
            point meta=explicit,draw=gray
        ] table [meta=C] {
            x y C
0 0 990
1 0 1
2 0 6
3 0 29
0 1 0
1 1 918
2 1 0
3 1 30
0 2 3
1 2 0
2 2 739
3 2 10
0 3 19
1 3 86
2 3 6
3 3 834
         };
    \end{axis}
%\draw[black,thick] (2.625,5.8) circle(0.5);
%\draw[red,dashed,thick] (3.675,5.8) circle(0.5);
\end{tikzpicture}}
    \\
    \multicolumn{2}{c}{(c) MLP}
    \end{tabular}
    \caption{Confusion matrices for classic techniques (BODMAS dataset)}
    \label{fig:CM_RF}
\end{figure}
    
When trained on~50 features, SVC achieves an accuracy of~0.9190, while our MLP classifier achieves~0.9482 accuracy.
The confusion matrices in Figures~\ref{fig:CM_RF}(b) and~(c) show us that both of these classifiers more often
misclassify \texttt{Sfone} as \texttt{Upatre}, as compared to any other misclassification.

\subsubsection{QR Code Experiments}

\begin{figure}[!htb]
    \centering
    \begin{tabular}{ccc}
    \includegraphics[width=35mm]{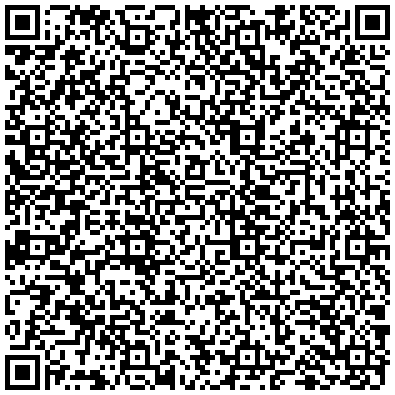}
    & &
    \includegraphics[width=35mm]{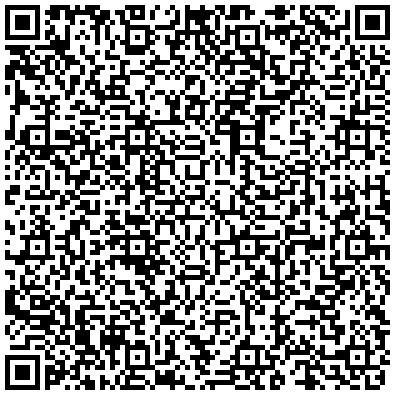}
    \\
    (a) Benign
    & &
    (b) \texttt{Sfone}
    \\ \\[-1.5ex]
    \includegraphics[width=35mm]{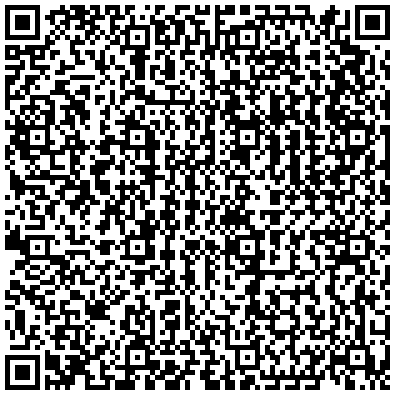}
    & &
    \includegraphics[width=35mm]{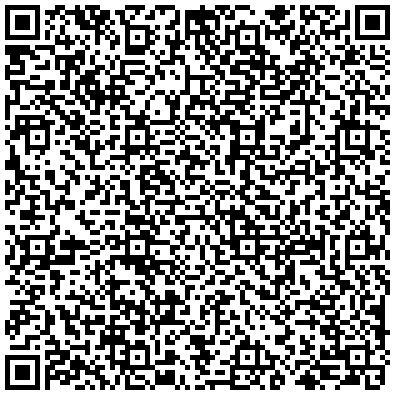}
    \\
    (c) \texttt{Upatre}
    & &
    (d) \texttt{Wacatac}
    \end{tabular}
    \caption{Examples of QR code representations (BODMAS)}
    \label{fig: BODMAS_BEINGN}
\end{figure}

To generate QR representations of the data, the top~50 features were used. 
Figures~\ref{fig: BODMAS_BEINGN}(a) through~(d) are examples of the 
benign, \texttt{Sfone}, \texttt{Upatre}, and \texttt{Wacatac} classes, respectively.

\begin{table}[!htb]
    \centering
    \caption{QR parameters for BODMAS dataset}
    \label{table:QR Parameters for BODMAS}
    \begin{tabular}{c|c}
    \toprule
    Parameter & Value \\ \midrule
    \texttt{version} & 1 \\
    \texttt{error\_correction} & \texttt{ERROR\_CORRECT\_L}\\
    \texttt{box\_size} & 5 \\
    \texttt{border} & 1 \\ \bottomrule
    \end{tabular}
\end{table}

The parameters used to generate these QR are given in 
Table~\ref{table:QR Parameters for BODMAS}.
The generated images are of size~$395 \times 395$ pixels. 
These were resized to~$128 \times 128$ images for use as input to our CNN. 

To train our CNN model, the~18,524 samples were split 80:20 to train:test. 
% The number of samples is not much different than in the previous case
The validation split was not done in this case due to the smaller number of samples available. 
The loss and accuracy graphs for this CNN model
are shown in the following Figure~\ref{fig:BODMAS_GRAPH},
where we see some indications of overfitting.

\begin{figure}[!htb]
    \centering
    \includegraphics[width=120mm]{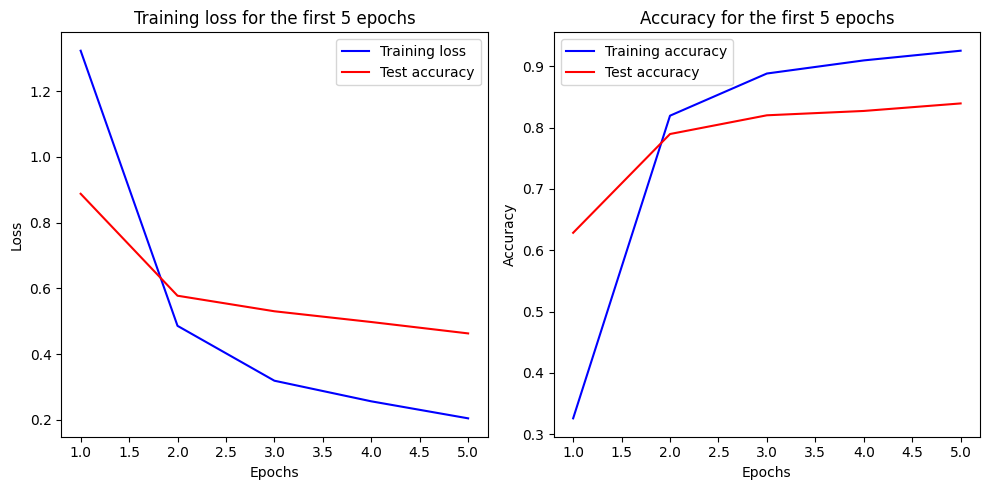}
    \caption{QR-CNN accuracy and loss graphs for BODMAS (50 features)}
    \label{fig:BODMAS_GRAPH}
\end{figure}

The test accuracy achieved for the CNN on the QR image representation was~0.8271 
for this multiclass problem. Note that this accuracy is less than we achieved with
each of our three classic ML techniques.

\begin{figure}[!htb]
    \centering
    \includegraphics[width=120mm]{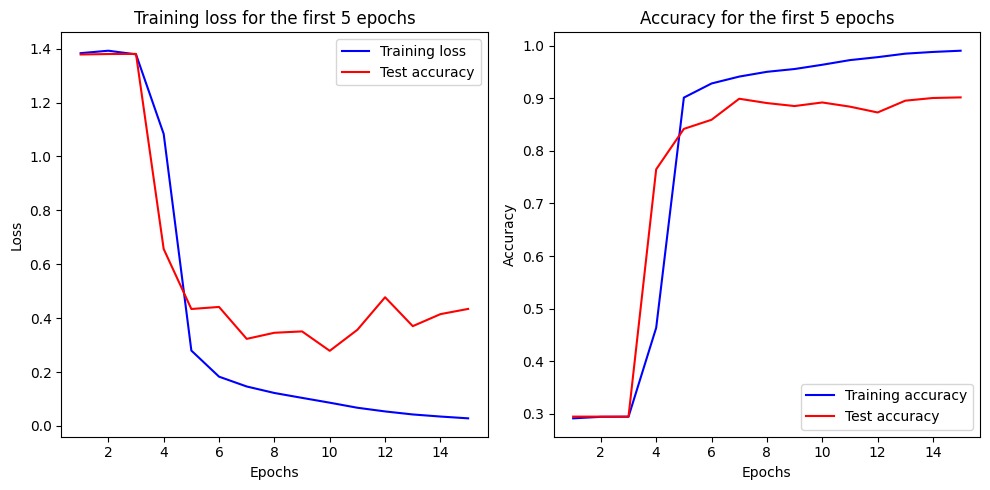}
    \caption{QR-CNN accuracy and loss graphs for BODMAS (150 features)}
    \label{fig:BODMAS_GRAPH_150_Features}
\end{figure}

We repeated this experiment using~150 features.
The loss and accuracy graphs for this case 
are shown in Figure~\ref{fig:BODMAS_GRAPH_150_Features}.
In this case, there appears to be less overfitting, as compared to the model based on~50 features.

The test accuracy achieved for the CNN based on~150 features is~0.8971. 
This result improves significantly on the case where~50 features are considered, 
but it is still less than our best classic ML technique.

\subsubsection{Aztec Code Experiments}

For the Aztec code representations of the data, we follow the same procedure as was used for our 
QR code experiments, above.
Specifically, we experiment using~50 features, then we repeat the entire set of experiments based on~150 features.

\begin{figure}[!htb]
    \centering
    \begin{tabular}{ccc}
    \includegraphics[width=35mm]{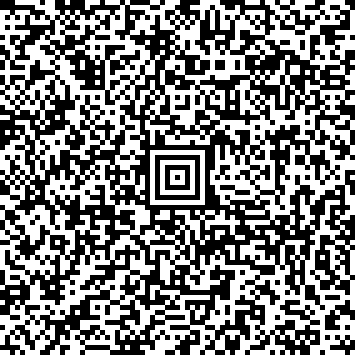}
    & &
    \includegraphics[width=35mm]{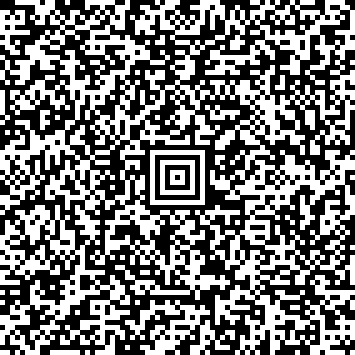}
    \\
    (a) Benign
    & &
    (b) \texttt{Sfone}
    \\ \\[-1.5ex]
    \includegraphics[width=35mm]{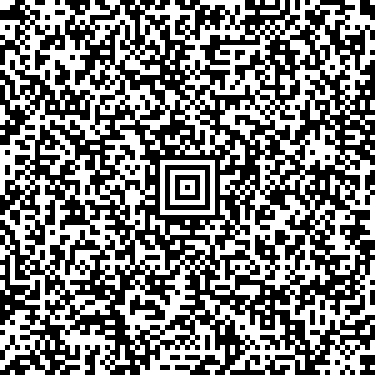}
    & &
    \includegraphics[width=35mm]{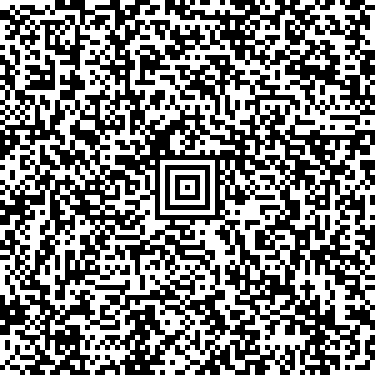}
    \\
    (c) \texttt{Upatre}
    & &
    (d) \texttt{Wacatac}
    \end{tabular}
    \caption{Examples of Aztec code representations (BODMAS)}
    \label{fig: BODMAS_BEINGN_AZTEC}
\end{figure}

Figures~\ref{fig: BODMAS_BEINGN_AZTEC}(a) through~(d) are representative examples of Aztec codes,
based on~50 features, for the benign, \texttt{Sfone}, \texttt{Upatre}, and \texttt{Wacatac} classes, respectively.
As above, the only parameter used fir these Aztec codes was~$\texttt{module\_size} = 5$.

The generated Aztec images are of size~$375 \times 375$ pixels. These are directly used as 
input to our CNN architecture, which is described in Section~\ref{chap:METHOD}. Note that no resizing is necessary.
Also, when training, we use an 80:20 split for training and testing. As with the QR code case,
a validation split was not used in this case. 

\begin{figure}[!htb]
    \centering
    \includegraphics[width=120mm]{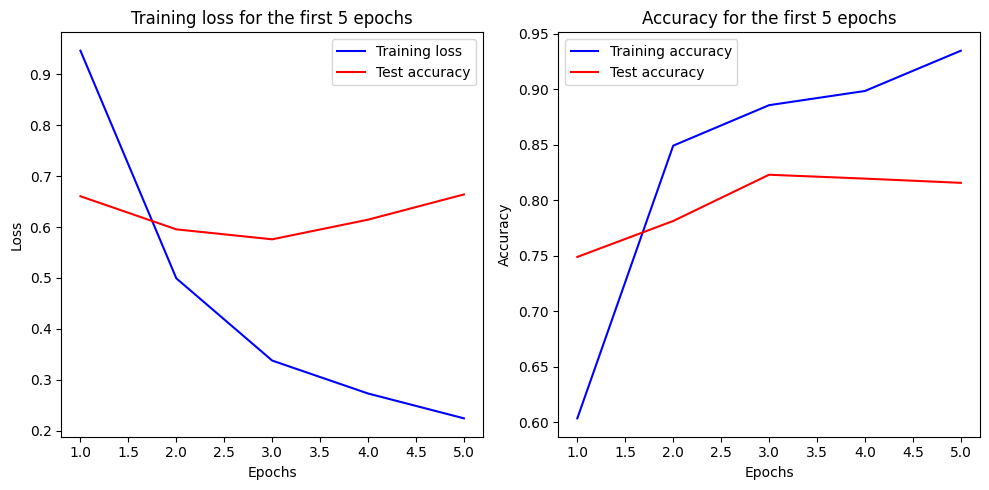}
    \caption{Aztec-CNN accuracy and loss graphs for BODMAS (50 features)}
    \label{fig:BODMAS_GRAPH_AZTEC_50}
\end{figure}

The loss and accuracy graphs for this case are shown in Figure~\ref{fig:BODMAS_GRAPH_AZTEC_50}.
It is clear that the model starts overfitting from epoch three. 
The test accuracy achieved for the CNN on the Aztec image representation was~0.7821 
for this multiclass problem.

We repeat the experiment above using Aztec images generated from~150 features.
All parameters are the same for this case as for~50 features case.
The loss and accuracy graphs for this experiment
are shown in the following Figure~\ref{fig:BODMAS_GRAPH_AZTEC_150}.
It is again clearly visible that the model is significantly overfitting the data. 
The test accuracy achieved for the CNN on the Aztec image representation 
based on~150 features is~0.8344.

\begin{figure}[!htb]
    \centering
    \includegraphics[width=120mm]{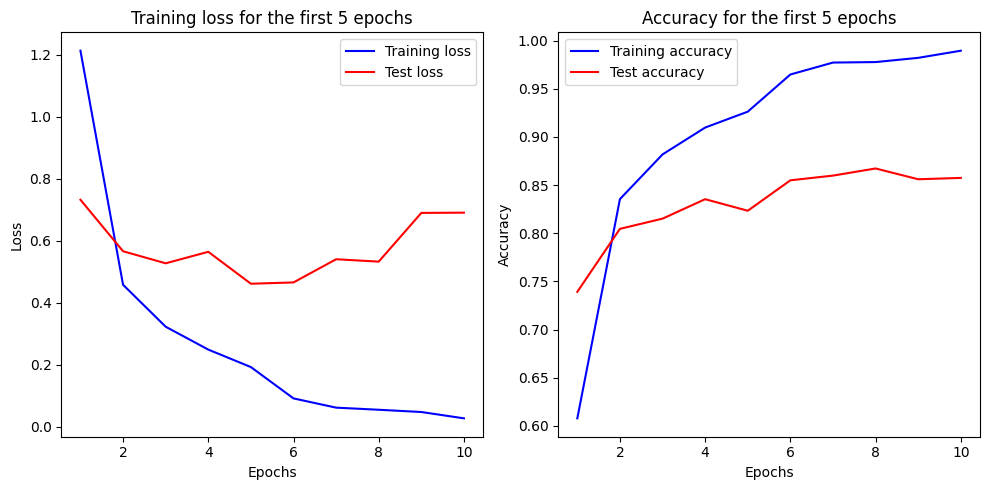}
    \caption{Aztec-CNN accuracy and loss graphs for BODMAS (150 features)}
    \label{fig:BODMAS_GRAPH_AZTEC_150}
\end{figure}

\subsection{Discussion}

Tables~\ref{table:classic model parameters} and~\ref{table:CNN model parameters} 
in Appendix~\ref{appendix_b}
list the hyperparameters tested (via grid search) 
for our classic machine learning and CNN experiments, respectively.
In these tables, we have also listed the accuracies obtained for each case.
Note that the accuracies in these tables are marginally better than the accuracies given 
in Sections~\ref{sect:resCIC} and~\ref{sect:resBOD} above, as here we have considered 
early stopping.

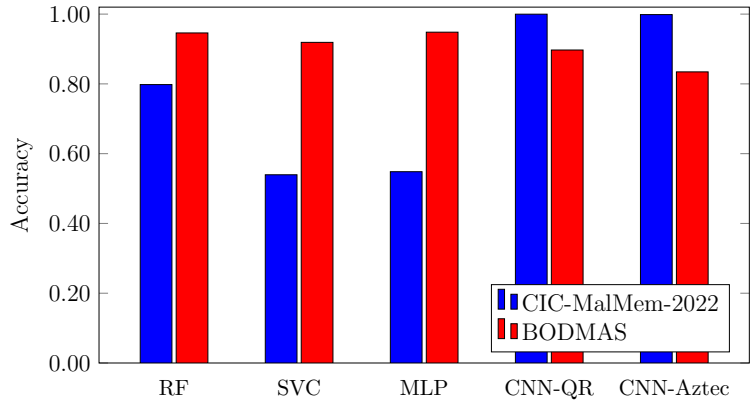
\begin{figure}[!htb]
    \centering
    \begin{tikzpicture}[scale=0.8, every node/.style={scale=0.95}]
\pgfkeys{/pgf/number format/.cd,1000 sep={}}
\begin{axis}[%bar shift=0pt,
        width  = 0.8*\textwidth,
        height = 7.5cm,
        ymin=0.0,ymax=1.02,
        ytick={0.0, 0.2, 0.4, 0.6, 0.8, 1.0},
        major x tick style = transparent,
        ybar=5*\pgflinewidth,
        bar width=15.0pt,
%        ymajorgrids = true,
%        xlabel = {Learning technique},
        ylabel = {Accuracy},
        symbolic x coords={RF, SVC, MLP, CNN-QR, CNN-Aztec},
        xticklabels={RF, SVC, MLP, CNN-QR, CNN-Aztec},
	y tick label style={
    		/pgf/number format/.cd,
   		fixed,
   		fixed zerofill,
%		sep=,
    		precision=2},
%	yticklabel pos=right,
        xtick = data,
        x tick label style={
		font=\small,
%		anchor=north east,
%		inner sep=0mm
		},
%		font=\small},
%        scaled y ticks = false,
	%%%%% numbers on bars and rotated
%        nodes near coords,
%        every node near coord/.append style={rotate=90, 
%        								   anchor=west,
%								   %font=\footnotesize,
%								   /pgf/number format/.cd,
%								   fixed,
%								   fixed zerofill,
%%								   sep=,
%								   precision=4},
        %%%%%
%        enlarge x limits=0.03,
%        enlarge x limits=0.06,
        enlarge x limits=0.15,
        legend cell align=left,
        legend pos=south east,
%        legend style={
%%                at={(1,1.05)},
%%                anchor=south east,
%%	        nodes={rotate=90},%%%%% rotate text in legend
%%                at={(0.125,0)},
%%                at={(0.125,0)},
%%                at={(0.8775,0)},
%                at={(0.82,0.56)},
%                anchor=south,
%                column sep=1ex
%        },
]
\addplot [fill=blue,opacity=1.00]
coordinates {
(RF, 0.7980)
(SVC, 0.5395)
(MLP, 0.5483)
(CNN-QR, 0.9998)
(CNN-Aztec, 0.9986)
};
\addlegendentry{CIC-MalMem-2022}
\addplot [fill=red,opacity=1.00]
coordinates {
(RF, 0.9460)
(SVC, 0.9190)
(MLP, 0.9482)
(CNN-QR, 0.8971)
(CNN-Aztec, 0.8344)
};
\addlegendentry{BODMAS}
\end{axis}
\end{tikzpicture}
    \caption{Accuracy comparison graph}
    \label{fig:Comparison_graph}
\end{figure}

The accuracies for the various models tested over the two datasets
are summarized in the form of a bar graph in Figure~\ref{fig:Comparison_graph}.
Note that for the CNN experiments on the BODMAS dataset, we have used the 
``150 Features'' results from Table~\ref{table:CNN model parameters},
 which were better than the ``50 Features'' results for both the QR and Aztec codes.

From Figure~\ref{fig:Comparison_graph}, we observe that the QR and Aztec codes far outperform 
classic techniques on the obfuscated CIC-MalMem-2022 dataset. However, for the BODMAS dataset, 
the situation is very different, with all three of the classic learning techniques tested 
outperforming our CNN architectures, regardless of whether the CNN was 
trained on QR or Aztec codes.

\section{Conclusion and Future Work}\label{chap:CONCLUSION}

In recent years, malware detection and classification based on image 
analysis has received considerable attention in the literature.
The method used to construct images from malware can have a
major impact on the success of such techniques, yet this aspect
of image-based malware analysis has received relatively little attention.

In this paper, we provided an empirical analysis of the utility
of QR and Aztec codes when used to provide image representions
of features extracted from malware. We compared 
CNN models trained on these code images to learning techniques
trained directly on the features, using two distinct datasets. Based on these experiments,
we found that for the CIC-MalMem-2022
dataset---which consists of dynamic features
extracted from obfuscated malware---the QR and Aztec code results were remarkably good.
On the other hand, for the more typical malware samples in the BODMAS dataset---which 
consists of static features---our QR-CNN and Aztec-CNN results did not improve
on other, non-image learning approaches. That is, classic ML techniques
trained on non-image features performed better on the BODMAS than our
more complex QR and Aztec image-based techniques.

There are many possible avenues for future work. Perhaps most urgently, we would like to
understand why the QR and Aztec codes perform extremely well on the CIC-MalMem-2022 dataset,
yet yielded inferior results on the BODMAS dataset. There are at least three possible reasons for
this discrepancy. 
\begin{itemize}
\item The CIC-MalMem-2022 dataset is derived from obfuscated malware,
while BODMAS is not. Perhaps code-based images are superior on 
more challenging cases, such as obfuscated malware. 
\item The features in the CIC-MalMem-2022 dataset were determined via dynamic analysis, 
while the BODMAS features are based on static analysis.
It is conceivable that the dynamic features are more informative, and that CNNs
trained on code-based images are better able to take advantage of this additional information.
\item The CNNs that we trained on the BODMAS dataset showed clear signs of overfitting.
Perhaps CNNs would achieve stronger results on this dataset if we reduce this
overfitting. Various techniques are available that can often mitigate overfitting
in CNNs. For example, cutout regularization~\cite{cutout}, which is somewhat analogous to 
the popular dropout regularization~\cite{dropout} used in other types of neural networks,
would be worth testing.
\end{itemize}

The BODMAS dataset includes accurate timestamps, and hence it is ideal for the
study of concept drift~\cite{drift}, which refers to the need to update models when the
underlying data has changed. Such ``drift'' is common in malware, where
families evolve as new features are added---existing malware may be adapted for
other purposes, new obfuscation techniques may be applied, and so on.
If improved results can be obtained for CNNs trained on the BODMAS dataset
based on QR or Aztec codes, then testing the robustness of such models 
under concept drift would be interesting.

Additional tests of QR and Aztec code representations on other malware datasets, as well as other 
classification problems involving inherently non-image data, would be interesting. Such experiments would
enable us to determine the relative strengths and weaknesses of code-based data 
representations in the realm of machine learning.

\bibliographystyle{plain}
\bibliography{references.bib}

%\appendix
%\makeatletter
%\renewcommand\section{\@startsection {section}{1}{\z@}%
%                                     {-3.5ex \@plus -1ex \@minus -.2ex}%
%                                     {2.3ex \@plus.2ex}%
%                                     {\noindent\normalfont\Large\bfseries Appendix }}
%\makeatother
%
%\section{}\label{appendix_a}

\titleformat{\section}{\normalfont\large\bfseries}{}{0em}{#1\ \thesection}
\setcounter{section}{0}
\renewcommand{\thesection}{\Alph{section}}
\renewcommand{\thesubsection}{A.\arabic{subsection}}
\setcounter{table}{0}
\renewcommand{\thetable}{A.\arabic{table}}
\setcounter{figure}{0}
\renewcommand{\thefigure}{A.\arabic{figure}}
\section{Appendix}\label{appendix_a}

In Table~\ref{tab:longtable}, we provide a complete list of the~55 
features that appear in the CIC-MalMem-2022 dataset. 
These features form the basis of experiments discussed
in Section~\ref{sect:resCIC} of this paper,
and the top~10 most informative of these~55 features 
are listed in Table~\ref{table:Top 10 features_CIC},

Recall that the features listed in Table~\ref{tab:longtable}
are derived from memory dumps of selected malware samples. 
For additional information on these
features, see~\cite{malmem}.

\begin{table}
\caption{List of features for CIC-MalMem-2022 dataset}\label{tab:longtable}
\centering
\adjustbox{scale=0.7}{
\begin{tabular}{c|ccc|c}
%\toprule
\cmidrule[\heavyrulewidth]{1-2} \cmidrule[\heavyrulewidth]{4-5}
Index & Feature & \phantom{NN} & Index & Feature \\ \cmidrule{1-2} \cmidrule{4-5}
\zz1 & \texttt{pslist.nproc} & & 29 & \texttt{malfind.protection} \\
\zz2 & \texttt{pslist.nppid} & & 30 & \texttt{malfind.uniqueInjections} \\
\zz3 & \texttt{pslist.avg\_threads} & & 31 & \texttt{psxview.not\_in\_pslist} \\
\zz4 & \texttt{pslist.nprocs64bit} & & 32 & \texttt{psxview.not\_in\_eprocess\_pool} \\
\zz5 & \texttt{pslist.avg\_handlers} & & 33 & \texttt{psxview.not\_in\_ethread\_pool} \\
\zz6 & \texttt{dlllist.ndlls} & & 34 & \texttt{psxview.not\_in\_pspcid\_list} \\
\zz7 & \texttt{dlllist.avg\_dlls\_per\_proc} & & 35 & \texttt{psxview.not\_in\_csrss\_handles} \\
\zz8 & \texttt{handles.nhandles} & & 36 & \texttt{psxview.not\_in\_session} \\
\zz9 & \texttt{handles.avg\_handles\_per\_proc} & & 37 & \texttt{psxview.not\_in\_deskthrd} \\
10 & \texttt{handles.nport} & & 38 & \texttt{psxview.not\_in\_pslist\_false\_avg} \\
11 & \texttt{handles.nfile} & & 39 & \texttt{psxview.not\_in\_eprocess\_pool\_false\_avg} \\
12 & \texttt{handles.nevent} & & 40 & \texttt{psxview.not\_in\_ethread\_pool\_false\_avg} \\
13 & \texttt{handles.ndesktop} & & 41 & \texttt{psxview.not\_in\_pspcid\_list\_false\_avg} \\
14 & \texttt{handles.nkey} & & 42 & \texttt{psxview.not\_in\_csrss\_handles\_false\_avg} \\
15 & \texttt{handles.nthread} & & 43 & \texttt{psxview.not\_in\_session\_false\_avg} \\
16 & \texttt{handles.ndirectory} & & 44 & \texttt{psxview.not\_in\_deskthrd\_false\_avg} \\
17 & \texttt{handles.nsemaphore} & & 45 & \texttt{modules.nmodules} \\
18 & \texttt{handles.ntimer} & & 46 & \texttt{svcscan.nservices} \\
19 & \texttt{handles.nsection} & & 47 & \texttt{svcscan.kernel\_drivers} \\
20 & \texttt{handles.nmutant} & & 48 & \texttt{svcscan.fs\_drivers} \\
21 & \texttt{ldrmodules.not\_in\_load} & & 49 & \texttt{svcscan.process\_services} \\
22 & \texttt{ldrmodules.not\_in\_init} & & 50 & \texttt{svcscan.shared\_process\_services} \\
23 & \texttt{ldrmodules.not\_in\_mem} & & 51 & \texttt{svcscan.interactive\_process\_services} \\
24 & \texttt{ldrmodules.not\_in\_load\_avg} & & 52 & \texttt{svcscan.nactive} \\
25 & \texttt{ldrmodules.not\_in\_init\_avg} & & 53 & \texttt{callbacks.ncallbacks} \\
26 & \texttt{ldrmodules.not\_in\_mem\_avg} & & 54 & \texttt{callbacks.nanonymous} \\
27 & \texttt{malfind.ninjections} & & 55 & \texttt{callbacks.ngeneric} \\ 
28 & \texttt{malfind.commitCharge} & & --- & --- \\
\cmidrule[\heavyrulewidth]{1-2} \cmidrule[\heavyrulewidth]{4-5}
\end{tabular}
}
\end{table}

%\clearpage

\renewcommand{\thesubsection}{B.\arabic{subsection}}
\setcounter{table}{0}
\renewcommand{\thetable}{B.\arabic{table}}
\setcounter{figure}{0}
\renewcommand{\thefigure}{B.\arabic{figure}}
\section{Appendix}\label{appendix_b}

In Tables~\ref{table:classic model parameters} and~\ref{table:CNN model parameters}
we list the hyperparameters tested (via grid search) 
for the classic techniques and our CNN models, respectively.
Note that for each model, the selected values are given in boldface.

\begin{table}[!htb]
\centering
\caption{Hyperparameters tested for classic techniques (selected values in boldface)}
\label{table:classic model parameters}
\adjustbox{scale=0.775}{
\begin{tabular}{cc|cc|c}
\toprule
Dataset & Classifier
 & Hyperparameters & Tested values & Accuracy \\
\midrule
\multirow{3}{*}{CIC-MalMem-2022} &
\multirow{3}{*}{RF}  & \texttt{n\_esitmators} & $\left[100, 500, \textbf{1000}\right]$ & \multirow{3}{*}{0.7980} \\
& & \texttt{criterion} & $\left [\textbf{gini}, \mbox{entropy}\right]$ & \\
& & Number of features & $\left [\textbf{10}, 55\right]$ & \\
\midrule
\multirow{3}{*}{CIC-MalMem-2022} &
\multirow{3}{*}{SVC}  & \texttt{kernel} & $\left [\textbf{rbf}, \mbox{poly}\right]$ & \multirow{3}{*}{0.5395} \\
& & \texttt{gamma} & $\left [\mbox{auto}, \textbf{scale}\right]$ & \\
& & Number of features & $\left [\textbf{50}, 150\right]$ & \\
\midrule
\multirow{4}{*}{CIC-MalMem-2022} &
\multirow{4}{*}{MLP}  & \texttt{hidden\_layer\_sizes} & $\left[\textbf{(10,10,10)}, (15,10,10)\right]$ & \multirow{4}{*}{0.5483} \\
& & \texttt{solver} & $\left [\mbox{sgd}, \textbf{Adam}\right]$ & \\
& & \texttt{max\_iter} & $\left [50, 100, \textbf{200}\right]$ & \\
& & Number of features & $\left [10, \textbf{55}\right]$ & \\
\midrule
\multirow{3}{*}{BODMAS} &
\multirow{3}{*}{RF}  & \texttt{n\_esitmators} & $\left[100, 500, \textbf{1000}\right]$ & \multirow{3}{*}{0.9460} \\
& & \texttt{criterion} & $\left [\textbf{gini}, \mbox{entropy}\right]$ & \\
& & Number of features & $\left [\textbf{50}, 150\right]$ & \\
\midrule
\multirow{3}{*}{BODMAS} &
\multirow{3}{*}{SVC}  & \texttt{kernel} & $\left[\textbf{rbf}, \mbox{poly}\right]$ & \multirow{3}{*}{0.9190} \\
& & \texttt{gamma} & $\left [\mbox{auto}, \textbf{scale}\right]$ & \\
& & Number of features & $\left [\textbf{50}, 150\right]$ & \\
\midrule
\multirow{4}{*}{BODMAS} &
\multirow{4}{*}{MLP}  & \texttt{hidden\_layer\_sizes} & $\left[\textbf{(10,10,10)}, (15,10,10)\right]$ & \multirow{4}{*}{0.9482} \\
& & \texttt{solver} & $\left [\mbox{sgd}, \textbf{Adam}\right]$ & \\
& & \texttt{max\_iter} & $\left [50, 100, \textbf{200}\right]$ & \\
& & Number of features & $\left [\textbf{50}, 150\right]$ & \\
\bottomrule
\end{tabular}
}
\end{table}

\begin{table}[!htb]
\centering
\caption{Hyperparameters tested for CNNs (selected values in boldface)}
\label{table:CNN model parameters}
\adjustbox{scale=0.775}{
\begin{tabular}{cc|cc|cc}
\toprule
\multirow{2}{*}{Dataset} & \multirow{2}{*}{Code}
 & \multirow{2}{*}{Hyperparameters} & \multirow{2}{*}{Tested values} & \multicolumn{2}{c}{Accuracy} \\
& & & & Train & Test \\
\midrule
\multirow{5}{*}{CIC-MalMem-2022} &
\multirow{5}{*}{QR}  & \texttt{learning\_rate} & $\left[\textbf{0.001}, 0.0001\right]$ & \multirow{5}{*}{0.9996} & \multirow{5}{*}{0.9998} \\
& & \texttt{batch\_size} & $\left [32, \textbf{64}, 128\right]$ & & \\
& & Epochs & $\left [\textbf{5}, 10, 20\right]$& & \\
& & Optimizer & $\left[\textbf{Adam}, \mbox{RMSProp}\right]$ & & \\
& & \texttt{image\_dim} & $\left[\textbf{128}, 175\right]$ & & \\
\midrule
\multirow{5}{*}{CIC-MalMem-2022} & 
\multirow{5}{*}{Aztec} & \texttt{learning\_rate} & $\left[\textbf{0.001}, 0.0001\right]$ & \multirow{5}{*}{0.9998} & \multirow{5}{*}{0.9986} \\
& & \texttt{batch\_size} & $\left [32, \textbf{64}, 128\right]$ & & \\
& & Epochs & $\left [\textbf{5}, 10, 20\right]$ & & \\
& & Optimizer & $\left[\textbf{Adam}, \mbox{RMSProp}\right]$ & & \\
& & \texttt{image\_dim} & $\left[\textbf{128}, 175\right]$ & & \\
\midrule
\multirow{5}{*}{BODMAS} &
\multirow{5}{*}{QR} & \texttt{learning\_rate} & $\left[\textbf{0.001}, 0.0001\right]$ & \multirow{5}{*}{0.9386} & \multirow{5}{*}{0.8271} \\
\multirow{5}{*}{(50 Features)} & & \texttt{batch\_size} & $\left [32, \textbf{64}, 128\right]$ & & \\
& & Epochs & $\left [\textbf{5}, 10, 20\right]$ & & \\
& & Optimizer & $\left[\textbf{Adam}, \mbox{RMSProp}\right]$ & & \\
& & \texttt{image\_dim} & $\left[\textbf{128}, 256\right]$ & & \\
\midrule
\multirow{5}{*}{BODMAS} &
\multirow{5}{*}{Aztec} & \texttt{learning\_rate} & $\left[\textbf{0.001}, 0.0001\right]$ & \multirow{5}{*}{0.9288} & \multirow{5}{*}{0.7821} \\
\multirow{5}{*}{(50 Features)} & & \texttt{batch\_size} & $\left [32, \textbf{64}, 128 \right]$ & & \\
& & Epochs & $\left [\textbf{5},10,20\right]$ & & \\
& & Optimizer & $\left[\textbf{Adam}, \mbox{RMSProp}\right]$ & & \\
& & \texttt{image\_dim} & $\left[\textbf{128}, 256, 375\right]$ & & \\
\midrule
\multirow{4}{*}{BODMAS} &
\multirow{4}{*}{QR} & \texttt{learning\_rate} & $\left[\textbf{0.001}, 0.0001\right]$ & \multirow{4}{*}{0.9762} & \multirow{4}{*}{0.8971} \\
\multirow{4}{*}{(150 Features)} & & \texttt{batch\_size} & $\left[32, \textbf{64}, 128\right]$ & & \\
& & Epochs & $\left [\textbf{5}, 10, 20\right]$ & & \\
& & \texttt{image\_dim} & $\left[128, \textbf{256}, 512, 675\right]$ & & \\
\midrule
\multirow{5}{*}{BODMAS} &
\multirow{5}{*}{Aztec} & \texttt{learning\_rate} & $\left[\textbf{0.001}, 0.0001\right]$ & \multirow{5}{*}{0.9521} & \multirow{5}{*}{0.8344} \\
\multirow{5}{*}{(150 Features)} & & \texttt{batch\_size} & $\left [32, \textbf{64}, 128 \right]$ & & \\
& & Epochs & $\left [5, \textbf{10}, 20\right]$ & & \\
& & Optimizer & $\left[\textbf{NAdam}, \mbox{Adam}\right]$ & & \\
& & \texttt{image\_dim} & $\left[128, \textbf{256}, 512, 675\right]$ & & \\
\bottomrule
\end{tabular}
}
\end{table}

\end{document}